\documentclass[%
 reprint,
 amsmath,amssymb,
 aps,showkeys,https://www.overleaf.com/project/608bd335d666745eb22d6f68
]{revtex4-2}

\usepackage{xcolor}

\usepackage[ruled]{algorithm2e}
\usepackage{siunitx}
\usepackage{graphicx}
\usepackage{dcolumn}
\usepackage{bm}
\usepackage{hyperref}
\usepackage{multirow}
\DeclareMathOperator*{\argmax}{\arg\!\max}

\begin{document}

\title{Accelerating global parameter estimation of gravitational waves from Galactic binaries using a genetic algorithm and GPUs}

\author{Stefan H. Strub}
 \email{stefan.strub@erdw.ethz.ch}
\author{Luigi Ferraioli}%
\author{Cédric Schmelzbach}%
\author{Simon C. Stähler}%
\author{Domenico Giardini}%
\affiliation{%
Institute of Geophysics, ETH Zurich\\ Sonneggstrasse 5, 8092 Zurich, Switzerland
}%

\begin{abstract}

The Laser Interferometer Space Antenna (LISA) is a planned space-based gravitational wave telescope with the goal of measuring gravitational waves in the milli-Hertz frequency band, which is dominated by millions of Galactic binaries. While some of these binaries produce signals that are loud enough to stand out and be extracted, most of them blur into a confusion foreground. Current methods for analyzing the full frequency band recorded by LISA to extract as many Galactic binaries as possible and to obtain Bayesian posterior distributions for each of the signals are computationally expensive. We introduce a new approach to accelerate the extraction of the best fitting solutions for Galactic binaries across the entire frequency band from data with multiple overlapping signals. Furthermore, we use these best fitting solutions to omit the burn-in stage of a Markov chain Monte Carlo method and to take full advantage of GPU-accelerated signal simulation, allowing us to compute posterior distributions in 2 seconds per signal on a laptop-grade GPU.
\end{abstract}

\maketitle

\section{Introduction}

The detection of gravitational waves (GWs) by the LIGO detector in 2015 marked a significant breakthrough in astrophysics  \cite{abbott2019gwtc}. This achievement spurred the development of the Laser Interferometer Space Antenna (LISA), a space-based interferometric system capable of detecting low frequency GWs in the $\SI{[0.1, 100]} {\mathrm{mHz}}$ range, free from terrestrial seismic and anthropogenic noise sources \cite{amaro2017laser}. LISA is a L-class mission of the European Space Agency (ESA) and currently set for launch in 2037.

The primary sources in the LISA frequency band are tens of millions of Galactic binaries (GBs) emitting quasi-monochromatic gravitational waves. These sources are far from merging, allowing for their gravitational waves to be continuously measured during LISA's nominal 4 year operational time \cite{amaro2017laser}. It is estimated that tens of thousands of these overlapping signals are resolvable by an experiment of LISA's arm length, resolution and measurement duration, while the rest blurs into a galactic foreground noise. Accurately estimating the parameters of GBs provides valuable information for studying the dynamical evolution of binaries \cite{taam1980gravitational, willems2008probing, nelemans2010chemical, littenberg2019binary, piro2019inferring}.

Several methods have been proposed for extracting GB signals, including maximum likelihood estimate (MLE)  \cite{ bouffanais2016DE, zhang2021resolving, gao2023fast} and Bayesian approaches. MLE methods are used to find the best matching simulated signal to the data, while Bayesian methods provide a posterior distribution that describes the uncertainty of the source parameters. The most successful Bayesian approaches are Markov chain Monte Carlo (MCMC) based methods, such as blocked annealed Metropolis-Hastings (BAM) \cite{PhysRevD.75.043008, crowder2007genetic, littenberg2011detection}, an MCMC algorithm with simulated annealing, or the reversible jump Markov chain Monte Carlo (RJMCMC) \cite{littenberg2020global, littenberg2023prototype} method, which allows for varying parameter dimensions and thus variable numbers of GBs to construct the posterior distribution.

In our previous work \cite{strub2022}, we demonstrated that signal extraction can be divided into two parts for both isolated and overlapping signals in the frequency domain. The first part involves optimizing the GB parameters in order to achieve the best fit between the simulated signal and the available data. In the second part, Gaussian process regression \cite{Rasmussen2006} is used to model the log-likelihood function, which allows for the computation of the posterior distribution without the need to simulate the GW signal for each sample. In this paper, we extend the work to analyze the full galactic signal population from a simulated LISA data stream.

Furthermore, with recent advances in simulating a GB signal using GPUs, we swapped the Gaussian process regression modeling with directly computing the log-likelihood function using a GPU \cite{katz2022assessing}. For sampling, we use a Metropolis-Hastings algorithm with a proposal distribution independent of the current state of the chain. Therefore, we can make full use of calculating the log-likelihood for 10'000 signals in parallel and build the Markov chain in the next step. That way we are able to compute the posterior distribution of a single signal within only 1.8 seconds on a late 2018-released Quadro RTX 4000 Mobile GPU built inside a laptop. We demonstrate the benefit of such a speed up by solving for the GB of the LISA Data Challenge  (LDC)1-4, part of LDC1, which is also called Radler \cite{LDC}. This challenge encompasses a dataset containing instrument noise as well as 26 million GB signals. Additionally, the pipeline has also been tested on LDC2a, called Sangria, where the injected MBHBs are subtracted, resulting in a dataset comprising 30 million GBs along with instrument noise \cite{LDC}. 

In Section \ref{sec:bayes} we introduce Bayesian parameter estimation, and Section \ref{sec:full galaxy} provides a detailed description of the new pipeline. In Section \ref{sec:results} the performance of the pipeline is showcased through its successful handling of the LISA Data Challenges LDC1-4 and LDC2a. Lastly, Section \ref{sec:conclusion} discusses the performance of the pipeline and the potential for further pipeline development.

\section{Bayesian formulation for Signal extraction}
\label{sec:bayes}
Gravitational Waves are ripples in the fabric of spacetime caused by the acceleration of massive objects, such as merging black holes, neutron stars and white dwarfs. LISA is a planned space-based mission designed to detect these elusive signals with unprecedented precision. However, the expected LISA data, denoted as $d(t)$, will be contaminated by instrument noise and unresolved signals, making the extraction of the underlying gravitational wave signal, denoted as $s(t, \theta)$, a challenging task. To tackle this, Bayesian inference and data analysis techniques provide a powerful framework. For convenience, we will omit the notation for dependence on $t$ for the data  $d$ and the signals $s(\theta)$ in the following.

In Bayesian inference, we aim to infer the probability distribution of the parameters $\theta$ describing the gravitational wave signal $s(\theta)$ given the observed data $d$. This is done using Bayes' theorem, which relates the posterior distribution $p(\theta | d)$, the prior distribution $p(\theta)$, the likelihood $p(d | \theta)$, and the model evidence $p(d)$ as follows:

\begin{equation}
p \left( \theta | d \right) = \frac{p \left( d | \theta \right) p \left( \theta \right)}{p \left(d \right)}
\end{equation}

The posterior distribution $p(\theta | d)$ represents the updated probability distribution of the parameters $\theta$ after taking into account the measured data $d$. The prior distribution $p(\theta)$ incorporates any prior knowledge or assumptions about the parameters. The model evidence $p(d)$ is a normalization factor that ensures the posterior distribution integrates to unity, and it is independent of $\theta$, hence does not affect the relative probabilities.

In GW data analysis, the likelihood $p(d | \theta)$ quantifies the probability of measuring the data stream $d$ given the parameters $\theta$ of the gravitational wave signal. The log-likelihood is commonly used due to its mathematical convenience and is defined as:

\begin{equation}
\label{eq:log-likelihood}
\log p(d | \theta) = -\frac{1}{2} \langle d-s(\theta) | d-s(\theta) \rangle,
\end{equation}

where $\langle x(t) | y(t) \rangle$ is the scalar product between two time-domain signals $x(t)$ and $y(t)$, and it is defined as:

\begin{equation}
\label{eq:scalar}
\langle x(t) | y(t) \rangle = 4 \mathcal{R} \left( \int_0^\infty \frac{\tilde{x}(f) \tilde{y}^*(f)}{S(f)} \, df \right),
\end{equation}

Here, $\tilde{x}(f)$ marks the Fourier transform of $x(t)$, and $S(f)$ is the one-sided power spectral density of the noise, which characterizes the noise properties of the LISA detector. The noise is estimated and constantly updated during the search. The noise estimate for GB analysis is discussed in Section \ref{sec:noise local} and \ref{sec:noise global}.

To eliminate the laser noise in the LISA arms' laser measurements, time-delay-interferometry (TDI) will be employed, which combines the measurements into three observables: X, Y, and Z \cite{tinto1999cancellation, Armstrong_1999, estabrook2000time, dhurandhar2002algebraic, tinto2014time}. Consequently, the data $d$ and the signal $s(\theta)$ consist of TDI responses with multiple channels, and we write the inner product as the following sum

 \begin{equation}
 \langle  d-s\left(\theta \right) |  d-s\left(\theta \right) \rangle = \sum_{\alpha \in \mathcal{M}} \langle  d_\alpha-s_\alpha\left(\theta \right) |  d_\alpha-s_\alpha\left(\theta \right) \rangle 
 \end{equation}

Here, $\mathcal{M} = { X, Y, Z }$ represents the default TDI setting, or $\mathcal{M} = { A, E, T }$ where

\begin{equation}
    \begin{aligned}
 A &= \frac{1}{\sqrt{2}} \left( Z - X \right) \\
 E &= \frac{1}{\sqrt{6}} \left( X - 2Y + Z \right) \\
 T &= \frac{1}{\sqrt{3}} \left( X + Y + Z \right)
    \end{aligned}
 \end{equation}
 
are uncorrelated with respect to instrument noise \cite{vallisneri2005synthetic}. In this work we utilize $A$, $E$, and $T$. However, to save computational time, we consider only $A$ and $E$ for signals with frequencies $f < f_\ast / 2 = 1/(4 \pi L) \approx \SI{9.55}{mHz}$, as the contribution of the gravitational wave response for $T$ is suppressed \cite{littenberg2020global}. By setting the threshold at half the transfer frequency $f_\ast$, we adopt a more conservative approach.

\section{Extracting Galactic Binary signals in the full LISA frequency band}
\label{sec:full galaxy}

The simulation of a GW from a GB system involves eight parameters denoted as $\theta = \left\{\mathcal{A}, \lambda, \beta, f, \dot{f}, \iota, \phi_0, \psi\right\}$ \cite{cornish2007tests}. These parameters are utilized to model the GW signal, where $\mathcal{A}$ represents the amplitude, $\lambda$, and $\beta$ correspond to the sky coordinates in terms of ecliptic longitude and ecliptic latitude, respectively. The parameter $f$ represents the frequency of the GW, $\dot{f}$ denotes the first-order frequency derivative, $\iota$ represents the inclination angle, $\phi_0$ represents the initial phase, and $\psi$ corresponds to the polarization angle. In this study, we consider only the first-order frequency derivative and neglect higher-order frequency derivatives.

To obtain the MLE we can maximize the signal-to-noise ratio (SNR) defined as
\begin{equation}
\label{eq:SNR}
\rho =  \frac{\langle  d |  s\left(\theta^\prime \right) \rangle}{\sqrt{\langle  s\left(\theta^\prime \right) |  s\left(\theta^\prime \right) \rangle}} =  \frac{\langle  d |  s\left(\theta \right) \rangle}{\sqrt{\langle  s\left(\theta \right) |  s\left(\theta \right) \rangle}}.
\end{equation}
which is independent of $\mathcal{A}$ with $\theta^\prime = \theta \setminus \{ \mathcal{A}\}$ and obtain 

\begin{equation}
\label{eq:maxA}
\mathcal{A}_\text{max} =  \frac{\langle  d |  s\left(\theta^\prime \right) \rangle}{\langle  s\left(\theta^\prime \right) |  s\left(\theta^\prime \right) \rangle }
\end{equation}

analytically  \cite{strub2022}.

\subsection{Frequency segments}
Because fitting $>$10'000 signals globally is a currently untractable problem, we split the data into small segments in the frequency domain. In order to have a few signals in one segment while keeping the number of segments small for stability and efficiency, we determine the segment size to be double the size of the broadest signal expected for each frequency segment $B_{\textrm{segment}} (f) = 2 B_{\textrm{max}} (f)$. The width of a signal in the frequency domain is influenced by various factors contributing to signal broadening. These factors include the frequency change of the source itself, LISA's orbital motion around the sun, and LISA's cartwheel motion.

To obtain the widest expected broadening we multiply the highest frequency derivative with the observation time $B_{F} = \dot{f}_{\textrm{max}} T_{\textrm{obs}}$. Where $\dot{f}_{\textrm{max}}$ is determined by \cite{littenberg2020global}

\begin{equation}
\label{eq:f derivative}
    \dot{f} = \frac{96}{5} \pi^{8/3} \mathcal{M}_c^{5/3} f^{11/3}
\end{equation}
where $\mathcal{M}_c = \frac{(m_1 m_2)^{3/5}}{(m_1 + m_2)^{1/5}}$ is the chirp mass and  $f$ the frequency. For $\dot{f}_{\text{max}}$ the masses of the binary are set to the Chandrasekhar limit $m_1 = m_2 = \SI{1.4}{M_\odot}$ \cite{mazzali2007common}.

\begin{figure}[!ht]
\includegraphics[width=0.5\textwidth]{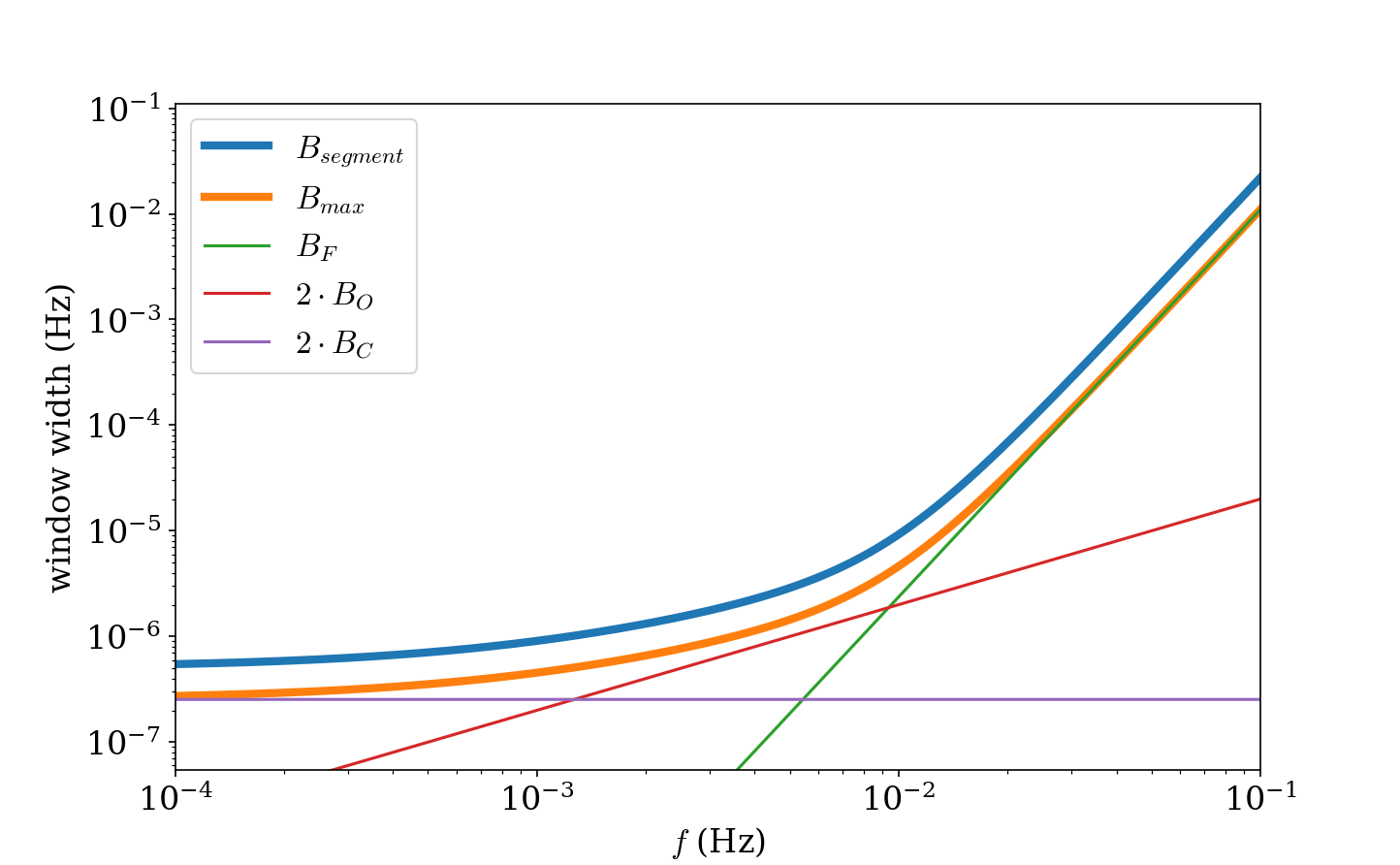}
\caption{\label{fig:bandwidth} Frequency segment widths to analyze GBs for $T_{obs} = \SI{2}{yr}$.}
\end{figure}

LISA's orbit around the sun and cartwheel motion smear the signal by $B_{\text{O}} = 10^{-4} f$ and $B_{\text{C}} = 4 \cdot \frac{1}{\SI{1}{yr}}$ respectively due to Doppler shift  \cite{Vallisneri_2009}. Since the smearing can increase or decrease the frequency, the resulting bandwidth is $2 B_{\text{O}}$ and $2 B_{\text{C}}$ respectively. As a result, the broadest signal expected has a width of $B_{\text{max}} = B_{\text{F}} + 2 B_{\text{O}} + 2 B_{\text{C}}$ which is shown in Figure \ref{fig:bandwidth}.

In Algorithm \ref{alg:frequency segments}, we outline the procedure for generating the list of frequency segments $B_\text{search}$ for a given global frequency interval. The lower bound of the frequency range, $f_\text{min} = \SI{0.3}{mHz}$, is chosen based on the absence of expected detectable GBs at frequencies lower than $\SI{0.3}{mHz}$. The upper bound, $f_\text{max} = f_\text{Nyquist}$, is determined by the sampling frequency where the Nyquist criterion states that the sampling frequency should be at least twice the maximum frequency of interest in order to accurately capture the signal \cite{shannon1949communication}.

\begin{algorithm}[H] 
\caption{Generating the list of frequency segments $B_\text{search}$ for searching GBs within the given ranges.}\label{alg:frequency segments}
$\textbf{Function} \ \textbf{\textit{segmenting}}(f_\text{min}, f_\text{max})$ \\
$B_\text{search} \gets \{ \ \}$ \\
$f \gets f_\text{min}$\\
\textbf{while} $f < f_\text{max} \textbf{ do}$\\
    \hspace*{1em} $f_\text{next} \gets f + 2B_{\text{max}}(f)$\\
    \hspace*{1em} $B_\text{search}$ append $ [f, f_{\text{next}}]$\\
    \hspace*{1em} $f \gets f_{\text{next}}$ \\
\textbf{return } $B_\text{search}$ 
\end{algorithm}

\subsection{Prior}
In Table \ref{tab:prior} we list the prior distribution $\Theta$ for all parameters. The frequency boundary is the padded frequency segment of interest $f_\text{segment} \in B_\text{search}$. The padding is half of the broadest signal expected $f_\text{padding} = \left(\max(f_\text{segment})-\min(f_\text{segment})\right)/4$ in case a signal is at the boundary of two neighboring segments like for example the yellow and grey signals at $\SI{4.226}{mHz}$ in Figure \ref{fig:4mHzdata}. For the upper bound of $\dot{f}$ we use $\dot{f}_\text{max}$ determined by \eqref{eq:f derivative}. Since we search for detached and interacting binaries the lower bound $\dot{f}$ is negative and is the same as in \cite{littenberg2020global}. The amplitude boundary is determined by a lower and upper bound SNR and is related to the amplitude by \cite{littenberg2020global}

\begin{equation}
\mathcal{A} \left( \rho \right) = 2 \rho \left( \frac{S\left(f\right)} {T_{obs} \, \sin^2\left(f / f_\ast \right)} \right)^{1/2}.
\end{equation}

\begin{table}[!ht]\footnotesize
\caption{Boundaries of the prior distribution $\Theta$.}
\begin{ruledtabular}
\begin{tabular}{ccc}
           Parameter &  Lower Bound & Upper Bound  \\ \hline
    $\sin \beta$  &     $-1$   &  1\\
  $\lambda$&      $-\pi$ & $\pi$\\
  $f$ & $\min (f_\text{segment}) - f_\text{padding}$  &  $\max(f_\text{segment}) + f_\text{padding} $\\
  $\dot{f}$ &  $ -5 \cdot 10^{-6} f^{13/3}$  & $1.02 \cdot 10^{-6} f^{11/3}$ \\
  $\log \mathcal{A}$ & $\log \mathcal{A} (\rho = 7)$ & $\log \mathcal{A} (\rho = 1000)$ \\
         $\cos \iota$&      $-1$  & 1\\
        $\phi_0$&          0  & 2$\pi$\\
         $\psi$&          0  & $\pi$ \\
\end{tabular}
\label{tab:prior}
\end{ruledtabular}
\end{table}

\subsection{Noise estimate within a frequency segment}
\label{sec:noise local}
For estimating the maximum likelihood of the GBs within a frequency segment the noise is estimated individually for each segment by calculating the periodogram \cite{auger1995improving, fulop2006algorithms}

\begin{equation}
    S_A(f) = \frac{2 |A(f)|^2}{N f_\textrm{sample}} 
\end{equation}

for each frequency window including the padding as determined for the prior listed in Table \ref{tab:prior}. $N$ marks the number of bins within the padded window and $f_\textrm{sample}$ represents the sampling frequency of the data $d$. In order to reduce the influence of loud signals within the window itself, the median of $S_A (f)$ is taken as the constant estimate for the full padded frequency segment. This brings a dynamic noise estimate during the search of signals which is updated after each found signal is subtracted from the data.  The estimate for other TDI variables $E, T$ is analog to the estimate of $A$.

\subsection{Galactic Binary search algorithm within a frequency segment}
In Algorithm \ref{alg:GB local} we present the GB search algorithm for given data $d_\textrm{analyze}$ to analyze on a given frequency segment $f_\textrm{segment} \in f_\textrm{search}$, which outputs a list $\tilde\theta_\textrm{in} = \{ \theta_\textrm{MLE,1}, \theta_\textrm{MLE,2}, ...\}$ of GB-parameters within the unpadded $f_\textrm{segment}$. Furthermore, $n_\text{signals}$ is the maximum number of signals per segment.

To save computational time, we limit the integral of the scalar product \eqref{eq:scalar} to the padded frequency segment. To obtain the MLE we use the differential evolution (DE) \cite{storn1997differential} algorithm and for the global optimization of all found signals within the unpadded region $\tilde\theta_\text{in}$ we use the Sequential Least Squares Programming (SLSP) method \cite{bonnans2006numerical}. Both methods are part of the SciPy library \cite{virtanen_scipy_2020}. The pipeline is set to search $n_\text{searches} = 3$ times for the same signal with varying initial parameters $\theta^\prime_{\textrm{init}} $ in case the search algorithm gets stuck at a local optimum.

Furthermore, we generalize the SNR to multiple signals $\tilde\theta = \{ \theta_\textrm{1}, \theta_\textrm{2}, ...\}$

\begin{equation}
\label{eq:SNR list}
\rho =  \frac{\left\langle  d |  \sum\limits_{\theta \in \tilde\theta}  s\left(\theta \right) \right\rangle}{\sqrt{\left\langle \sum\limits_{\theta \in \tilde\theta} s\left(\theta \right) | \sum\limits_{\theta \in \tilde\theta} s\left(\theta \right) \right\rangle}}.
\end{equation}

\begin{algorithm}[!ht]
\caption{The GB search algorithm within a frequency segment $f_\textrm{segment}$.}\label{alg:GB local}
$\textbf{Function} \ \textbf{\textit{local\_GB\_search}}( f_\text{segment}, n_\text{signals}, d_\text{analyze})$\\
$\tilde\theta_\textrm{found} \gets \{\,\}$\\
$\tilde\theta_\textrm{in} \gets \{\,\}$\\
$\tilde\theta_\textrm{out} \gets \{\,\}$\\
$d_\textrm{residual} \gets d_\textrm{analyze}$\\

\textbf{for $i$ in} $\{1,2,..., n_\textrm{signals}\}$  \textbf{do} \\
    \hspace*{1em} $\tilde\theta^\prime_\textrm{MLEs} \gets \{\,\}$\\
    
    \hspace*{1em}  \textbf{for $j$ in} $\{1,2,..., n_\textrm{searches}\}$  \textbf{do} \\
        \hspace*{2em} $\theta^\prime_{\textrm{init}} $ randomly drawn from prior\\
        \hspace*{2em} $\theta^\prime_\textrm{MLE} \gets  \argmax\limits_{\theta^\prime}  \rho (\theta^\prime, d_\textrm{residual}) $ using DE with $\theta^\prime_{\textrm{init}}$\\
        \hspace*{2em} \textbf{if} $\rho (\theta^\prime_\textrm{MLE}, d_\textrm{analyze}) \leq \rho_\textrm{threshold}-2$ \textbf{and} $j = 1 \textbf{ do}$\\
            \hspace*{3em} $\textbf{ return } \tilde\theta_\textrm{in}$\\
        \hspace*{2em} $\tilde\theta^\prime_\textrm{MLEs} \gets \tilde\theta^\prime_\textrm{MLEs} \cup \{\theta^\prime_\textrm{MLE}\}$\\
    \hspace*{1em} \textbf{end for}\\
    
    \hspace*{1em} $\theta^\prime_\textrm{MLE} \gets \argmax\limits_{\theta^\prime \in \tilde\theta^\prime_\textrm{MLEs}} \rho (\theta^\prime, d_\textrm{residual})$\\ 
    \hspace*{1em} \textbf{if} $\rho (\theta^\prime_\textrm{MLE}, d_\textrm{analyze}) \leq \rho_\textrm{threshold} \textbf{ do}$\\
        \hspace*{2em}$\textbf{return } \tilde\theta_\textrm{in}$\\
    \hspace*{1em} Compute $\mathcal{A}_\text{max}$ according to \eqref{eq:maxA} with $\theta^\prime_\textrm{MLE}$\\
    \hspace*{1em} $\theta_{\textrm{MLE}} \gets \theta^\prime_{\textrm{MLE}} \cup \{\mathcal{A}_\text{max}\}$\\
    \hspace*{1em} \textbf{if} $\theta^f_{\textrm{MLE}}$ in unpadded $f_\textrm{segment}$ \textbf{do}\\
    \hspace*{2em} $\tilde\theta_\textrm{in} \gets \tilde\theta_\textrm{in} \cup \{\theta_{\textrm{MLE}}\}$\\
    \hspace*{1em} \textbf{else do} \\
    \hspace*{2em} $\tilde\theta_\textrm{out} \gets \tilde\theta_\textrm{out} \cup \{\theta_{\textrm{MLE}}\}$\\
    \hspace*{1em} $d_\textrm{residual} \gets d_\textrm{analyze} - \sum\limits_{\theta \in \tilde\theta_\textrm{out}} s(\theta)$\\
    \hspace*{1em} $\tilde\theta_\textrm{in} \gets \argmax\limits_{\tilde\theta} \rho (\tilde\theta, r)$ using SLSP with $\tilde\theta_\textrm{in}$ as start\\
    \hspace*{1em} $\tilde\theta_\textrm{found} \gets \tilde\theta_\textrm{in} \cup \tilde\theta_\textrm{out}$\\
    \hspace*{1em} $d_\textrm{residual} \gets d_\textrm{analyze} - \sum\limits_{\theta \in \tilde\theta_\textrm{found}} s(\theta)$\\
\textbf{end for}\\
$\textbf{return } \tilde\theta_\textrm{in}$
\end{algorithm}

\begin{figure*}[!ht]
\includegraphics[width=1\textwidth]{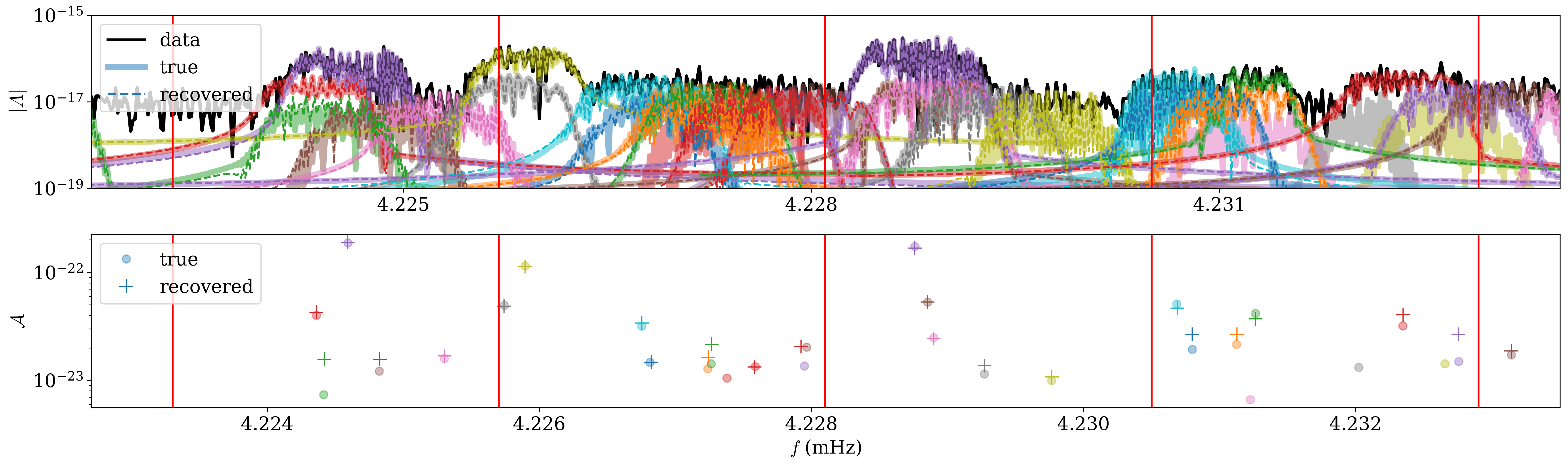}
\caption{\label{fig:4mHzdata} Displayed are the data, injected signals, and recovered signals of the Radler data challenge with $T_\text{obs} = \SI{2}{yr}$. The red lines are the boundaries of four adjacent frequency segments. The first plot illustrates the absolute value of the A TDI channel, while the second plot depicts the amplitude $\mathcal{A}$ across the frequency spectrum. The red lines mark the boundaries of the frequency segments. The plot is extended to the left and right by the padding of the segments at the borders.}
\end{figure*}

\subsection{Global GB search pipeline}
Segments that are not direct neighbors to each other can be analyzed in parallel as presented in Algorithm \ref{alg:GB global}. Therefore we enumerate all segments $B_\text{search}$ ordered by frequency and divide them into two groups of even and odd segments
\begin{equation*}
    \begin{aligned}
B_\text{even} &:= \{B_\text{search,i} \text{ : } i \in \{1, 2, ..., N_\text{segments}\} \text{ and i is even} \} \\
B_\text{odd} &:= \{B_\text{search,i} \text{ : } i \in \{1, 2, ..., N_\text{segments}\} \text{ and i is odd} \}
    \end{aligned}
\end{equation*}
where $N_\text{segments}$ is the number of segments in the list $B$. In Table \ref{tab:search parameters} we list the parameters: The list of frequency segments $f_\text{search}$, max number of signals per segment $n_\text{signals}$ and the data to analyze $d_\text{analyze}$. The analysis is conducted in three sequential runs, starting from the top line of the table, in order to cover all frequency segments within $B_\text{search}$. In the first run, we analyze all even segments $B_\text{even}$, allowing for a maximum of $n_\text{signals} = 3$ signals per segment. We assume that there are not more than 3 strong signals spilling into the padded regions for each segment.

Next, we proceed to analyze the odd segments in a similar manner. The found signals in the odd segments, denoted as $\tilde\theta_\text{odd}$, are subtracted from the original data $d$. Finally, we repeat the analysis of the even segments, now free from the influence of neighboring signals located in the neighboring odd segments. By subtracting the signals found in the odd segments and re-analyzing the even segments, we ensure that each segment of $B_\text{search}$ is analyzed independently without being affected by signals of neighboring signals.

\begin{algorithm}[!ht]
\caption{The search algorithm for multiple frequency segments $f_\textrm{search}$.}\label{alg:GB global}
$\textbf{Function} \ \textbf{\textit{global\_GB\_search}}(  f_\text{search}, n_\text{signals}, d_\text{analyze})$\\
$\tilde\theta \gets \{\,\}$\\
\textbf{for all} $f_\text{segment}$ \textbf{in} $f_\textrm{search}$  \textbf{do in parallel} \\
    \hspace*{1em} $\tilde\theta \gets \tilde\theta \cup \textbf{\textit{local\_GB\_search}}( f_\text{segment}, n_\text{signals}, d_\text{analyze})$\\
\textbf{end for}\\
$\textbf{return } \tilde\theta$
\end{algorithm}

The even segments where no signals in neighboring segments were detected and less than 3 signals were found are not analyzed a second time. Because the subtraction of the signals in odd windows did not influence these even segments and there is no need to repeat the search. For these segments, the found signals of the first even segments analysis are directly used for the catalog.

The LISA data will be a time-evolving data set with new data being constantly added. Therefore the found signals of previous runs can be used to speed up the analysis where $\tilde\theta_\text{initial}$ if $j = 1$ in Algorithm \ref{alg:GB local} is set to a signal within that frequency segments found in the previous run. Especially for signals $f > \SI{10}{mHz}$ and $T_\text{obs} > \SI{1}{yr}$ the success rate of \textbf{\textit{local\_GB\_search}} becomes small if $\tilde\theta_\text{initial}$ is randomly drawn from the prior. It is advantageous to use the found signals of a previous shorter data set analysis, for example, $T_\text{obs} = \SI{6}{months}$, as the initial value of the search algorithm.

The global solution is then $\tilde\theta_\text{recovered} = \tilde\theta_\text{even} \cup \tilde\theta_\text{odd}$ where $\tilde\theta_\text{even}$ is the solution of the third run. In Figure \ref{fig:4mHzdata} we show the solution $\tilde\theta_\text{recovered}$ for four neighboring segments at a region with multiple detectable and overlapping signals. We demonstrate with the pipeline a successful recovery rate of 25 out of 30 injected GBs. Among the 25 recovered signals, 24 of them correspond to individual injected signals, indicating a high level of accuracy in the recovery process. In addition, it is worth noting that the recovered signals at $f = \SI{4.22}{mHz}$ is a composite of two injected signals. However, the remaining unrecovered signal is characterized by low amplitudes $\mathcal{A}$.

\begin{table}[!ht]\footnotesize
\caption{Inputs and outputs of the search pipeline \textbf{\textit{global\_GB\_search}} across all frequency segments $B$ for given data $d$.}
\begin{ruledtabular}
\begin{tabular}{cccccc}
run & $f_\text{search}$ &  $n_\text{signals}$ & $d_\text{analyze}$ & output \\ \hline
1& \rule{0pt}{3ex}  $B_\text{even}$  & 3 & $d$ &  $\tilde\theta_\textrm{even}$\\
2&\rule{0pt}{3ex}  $B_\text{odd}$ & 10  & $d - \sum\limits_{\theta \in \tilde\theta_\textrm{even}} s(\theta)$ & $\tilde\theta_\textrm{odd}$ \\
3&\rule{0pt}{3ex}  $B_\text{even}$ & 10 & $d - \sum\limits_{\theta \in \tilde\theta_\textrm{odd}} s(\theta)$ & $\tilde\theta_\textrm{even}$\\
\end{tabular}
\label{tab:search parameters}
\end{ruledtabular}
\end{table}

\subsection{Global noise estimate}
\label{sec:noise global}
For the global noise estimate, we subtract each recovered signal $\tilde\theta_\text{recovered}$ from the data

\begin{equation}
    d_\text{residual} = d - \sum\limits_{\theta \in \tilde\theta_\text{recovered}} s(\theta).
\end{equation}
where $s(\theta)$ represents the signal corresponding to each MLE $\theta$. Furthermore, we proceed to estimate a smooth noise curve, denoted as $S_{A \text{,welch}}(f)$, across the entire frequency domain. This estimation is performed by applying Welch's method, utilizing 500 windows and a Hann window function \cite{blackman1958measurement}. Next, we address the remaining outlier peaks, mainly of unresolved signals, by implementing a smoothing procedure. We define a frequency window of 30 bins and adjust any values above the window's median to be twice the median value. This process is repeated by shifting the window by 15 frequency bins until the entire power spectral density (PSD) is smoothed. The result is denoted as $S_{A \text{,median}}(f)$.

To further enhance the smoothing effect, we utilize the Savitzky-Golay filter \cite{savitzky1964smoothing}. The filter is configured with an order of 1, and we apply two different window lengths depending on the frequency range. For observations with $T_\text{obs}$ equal to either 1 or 2 years, frequencies below $\SI{0.8}{mHz}$ are smoothed using a window length of 10, while frequencies above $\SI{0.8}{mHz}$ are smoothed using a window length of 70. In the case of $T_\text{obs} = \SI{0.5}{yr}$, frequencies below $\SI{0.8}{mHz}$ employ a window length of 10, and frequencies above $\SI{0.8}{mHz}$ are smoothed using a window length of 50.

Finally, to obtain a PSD estimate for each desired frequency bin, we spline interpolate the smoothed PSD, resulting in our estimate of the residual noise curve denoted as $S_{A \text{,residual}}(f)$.

\begin{figure}[!ht]
\includegraphics[width=0.5\textwidth]{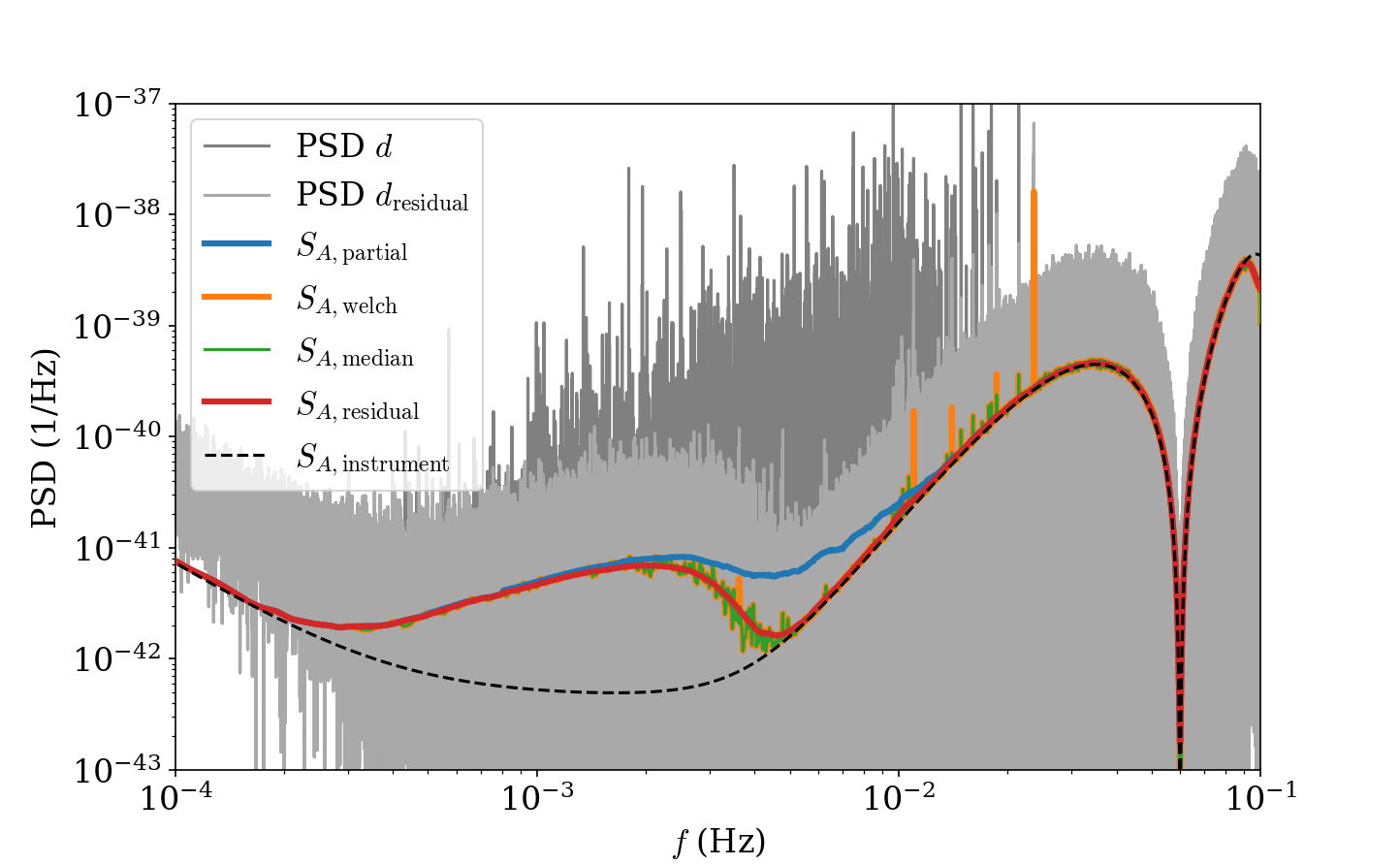}
\caption{Noise estimates and power spectrum density (PSD) of the TDI A channel of the $\SI{1}{yr}$ Sangria data set. $S_{A \text{,instrument}}$ is the noise PSD used for creating the data. The difference between $S_{A \text{, residual}}$ and the true PSD between $\SI{0.2}{mHz}$ and $\SI{5}{mHz}$ is due to the unresolved GBs which can be seen as red crosses in Figure \ref{fig:amplitude scatter}. It is expected, that most GBs in that frequency range are unresolvable and therefore merge into the galactic foreground noise.}\label{fig:noise}
\end{figure}

The noise estimates, depicted in Figure \ref{fig:noise}, exhibit a strong agreement with the instrument noise $S_{A \text{,instrument}}(f)$, except for the frequency range between $\SI{0.2}{mHz}$ and $\SI{5}{mHz}$. In this range, the unresolved background signals (GBs) merge into the galactic foreground noise, leading to deviations in the noise estimate. The noise of the other TDI channels $E$ and $T$ are computed the same way.

\subsection{GPU accelerated posterior distribution derivation}
\label{sec:posterior}

In order to derive the posterior distribution, we employ the Metropolis-Hastings Monte Carlo (MHMC) algorithm \cite{metropolis1953equation, hastings1970monte}. This algorithm suggests new parameters $\theta_\text{p}$ based on a proposal distribution $g(\theta_\text{p} | \theta_\text{c})$, which generally depends on the current state of the chain $\theta_\text{c}$. The proposed parameters are then accepted with probability 
\begin{equation}
P(\theta_\text{p},\theta_\text{c})=\min \left(1, \left[ {\frac {p(d \mid \theta_\text{p})}{p(d \mid \theta_\text{c})}}{\frac {g(\theta_\text{c} \mid \theta_\text{p})}{g(\theta_\text{p}\mid \theta_\text{c})}} \right] ^{\frac{1}{\mathcal{T}}} \right) 
\end{equation}
where $\mathcal{T}$ is the temperature for simulated annealing. 

Previously, \cite{strub2022} demonstrated that the MLE can be effectively utilized to accelerate the computation of the posterior distribution. The posterior distribution tends to be concentrated within a relatively compact region of the parameter space. As a result, it becomes unnecessary to sample beyond specific parameter space boundaries when employing Markov chain Monte Carlo (MCMC) methods to estimate the posterior. By identifying the reduced parameter space $\Theta_\text{reduced}$ where the posterior is concentrated, we can skip the burn-in phase typically required in MCMC sampling. Moreover, this approach allows for a proposal distribution $g(\theta_\text{p}) = g(\theta_\text{p} | \theta_\text{c})$ that is independent of the current state of the chain $\theta_\text{c}$. This is achieved by randomly drawing samples within $\Theta_\text{reduced}$. The independence from the chain's state enables the parallel computation of the log-likelihood for all samples in the first step, followed by the construction of the chain during the second step, where if the proposed sample $\theta_\text{p}$ is rejected the chain stays at the current sample $\theta_\text{c}$ as described in Algorithm \ref{alg:posterior}. This approach leverages the computational power of a GPU to rapidly compute the log-likelihood of 10'000 samples in parallel, facilitating a more efficient and rapid estimation of the posterior distribution.

To establish the reduced parameter space $\Sigma_\text{reduced}$ we use the inverse of the Fisher Information Matrix (FIM)
\begin{equation}
    F_{ij} = \langle \partial_i p(d \mid \theta_\text{MLE}) | \partial_j p(d \mid \theta_\text{MLE}) \rangle,
\end{equation}
where $\partial_i$ denotes the partial derivative with respect to the $i$\textsuperscript{th} component of the parameter vector $\theta$. To compute the derivatives of the FIM, the second-order forward finite difference method is employed with a step size of $10^{-9}$ times the search space determined by the prior distribution $\Theta$. The estimated uncertainty vector is $\sigma = \sqrt{\text{diag}(F^{-1})}$.

In our investigation, we set the volume of the parameter space to $\Theta_\text{reduced} = [ \theta_\text{MLE} - 4 \sigma, \theta_\text{MLE} + 4 \sigma]$. It is sufficient to set the boundary for the frequency parameter to $\Theta_\text{reduced}^f = [ \theta_\text{MLE}^f - \sigma_f, \theta_\text{MLE}^f + \sigma_f]$ where $\sigma_f$ denotes the estimated uncertainty of the frequency. The frequency derivative parameter space is not reduced and spans the full prior $\Theta_\text{reduced}^{\dot{f}} = \Theta^{\dot{f}}$. Due to degeneracy, we neglect the distribution of the polarization and the initial phase and define a narrow search space for them. Hence, we set $\Theta^\psi = [\theta^\psi_{\mathrm{MLE}} - \frac{\pi}{1000}, \theta^\psi_{\text{MLE}} + \frac{\pi}{1000}]$ and $\Theta^{\phi_0} = [\theta^{\phi_0}_\mathrm{MLE} - \frac{2\pi}{1000}, \theta^{\phi_0}_\mathrm{MLE} + \frac{2\pi}{1000}]$.

Simulated annealing is useful to further speed up the computation of the posterior. As a start, we use uniform sampling in the reduced parameter space $\Theta_\text{reduced}$ as the proposal distribution with high a temperature. Next, we utilize the obtained posterior distribution as the new proposal distribution by employing multivariate kernel density estimation (KDE) techniques \cite{o2016fast, o2014reducing}. To address the challenge of high-dimensional KDE computations, we group parameters into 2-dimensional parameter pairs. Specifically, we group $\mathcal{A} - \iota$, $\lambda - \beta$, and $f - \dot{f}$ together and perform KDE on each pair. This allows us to overcome the computational limitations associated with kernel density estimation involving four or more parameters. By gradually lowering the temperature $\mathcal{T}$ during the simulated annealing process, we can achieve more refined and accurate estimations of the posterior distribution while maintaining computational efficiency.

The results presented in the next Section \ref{sec:results} are created with six different temperatures $\mathcal{T} = \{ 15, 10, 5, 3, 2, 1 \}$ and constant $n_\text{samples} = 10\,000$. The number of samples $n_\text{samples}$ could be changed for each temperature.

\begin{algorithm}[!ht]
\caption{The GPU accelerated posterior distribution algorithm.}\label{alg:posterior}
$\textbf{Function} \ \textbf{\textit{posterior}}(\theta_\text{MLE}, d_\text{posterior}, \Theta_\text{reduced}, \tilde{\mathcal{T}}, n_\text{samples})$\\
$\Theta_\text{sample} \gets \Theta_\text{reduced}$\\
$\textbf{for } \mathcal{T} \textbf{ in } \tilde{\mathcal{T}} \textbf{ do}$\\
    \hspace*{1em} $\tilde\theta_\text{posterior} \gets \{\,\}$\\
    \hspace*{1em} $\tilde{L} \gets \{\,\}$\\
    \hspace*{1em} $\tilde\theta_\text{samples} \gets n_\text{samples} \text{ randomly drawn from } \Theta_\text{sample}$\\
    \hspace*{1em} $\textbf{for all } \theta \textbf{ in } \tilde\theta_\text{samples} \textbf{ do in parallel on GPU}$\\
        \hspace*{2em} $\tilde{L} \gets \tilde{L} \cup \{p(d_\text{posterior} \mid \theta)\}$\\
    \hspace*{1em} $\textbf{end for}$\\
    \hspace*{1em} $\theta_\text{c} \gets \tilde\theta_1$\\
    \hspace*{1em} $L_\text{current} \gets \tilde{L}_1$\\
    \hspace*{1em} $\textbf{for } i \textbf{ in } \{ 2, 3, ..., n_\text{samples} \} \textbf{ do}$\\
        \hspace*{2em} $\alpha \gets \min \left( 1, \left[ {\frac {L_i}{L_\text{current}}}{\frac {g(\theta_\text{c})}{g(\theta_i)}} \right] ^{\frac{1}{\mathcal{T}}} \right)$\\
        \hspace*{2em}with probability $\alpha \textbf{ do}$\\
    
            \hspace*{3em} $\theta_\text{c} = \tilde\theta_i$\\
            \hspace*{3em} $L_\text{current} = \tilde{L}_i$\\
        
        \hspace*{2em} $\tilde\theta_\text{posterior} \gets \tilde{\theta}_\text{posterior} \cup \{\theta_\text{c}\}$\\
    \hspace*{1em} $\textbf{end for}$\\
    \hspace*{1em} $\Theta_\text{sample} \gets \text{KDE}(\tilde\theta_\text{posterior})$\\
\textbf{end for}\\
\textbf{return } $\tilde\theta_\text{posterior}$\\
\end{algorithm}

The algorithm to compute the posterior distribution for a single signal $\theta_\text{MLE} \in \tilde\theta_\text{recovered}$ is presented in Algorithm \ref{alg:posterior}. The computation of the likelihood $p(d_\text{posterior} \mid \theta)$ on the GPU is based on \cite{michael_katz_2022_5867731} described in \cite{katz2022assessing}. The data for the input is

\begin{equation}
    d_\text{posterior} = d - \sum\limits_{\theta \in \tilde\theta_\text{recovered}} s(\theta) + s(\theta_\text{MLE}).
\end{equation}

The reduced parameter space $\Theta_\text{reduced}$ is determined with $\theta_\text{MLE}$ and $d_\text{posterior}$ as described above.

The resulting posterior distribution is the posterior given the overlapping MLEs $\tilde\theta_\text{overlap} \subset \tilde\theta_\text{gobal}$

\begin{equation}
    p(\theta_\text{MLE} | d_\text{posterior}) = p(\theta | d, \tilde\theta_\text{overlap} ) 
\end{equation}

which has a narrower posterior distribution than the marginalized posterior

\begin{equation}
    p(\theta_\text{MLE} | d) = \int p(\theta, \tilde\theta_\text{overlap} | d) p(\tilde\theta_\text{overlap}) \, \mathrm{ d} \tilde\theta_\text{overlap}.
\end{equation}

Overlapping signals lead to a joint posterior distribution. To approximate the marginalized posterior for such cases, one approach is to increase the estimated noise by computing the noise of the partial residual $S_{A \text{,partial}}(f)$. This is achieved by subtracting the found signals only partially from the original data, leaving some residual signal components in the data

\begin{equation}
    d_\text{partial} = d - s_\text{partial} \sum\limits_{\theta \in \tilde\theta_\text{recovered}} s(\theta).
\end{equation}
where $s_\text{partial} \in [0,1]$ is a scaling factor which we set to $s_\text{partial} = 0.7$. By analyzing this partial residual, one can obtain an approximation of the marginalized posterior distribution that takes into account the presence of overlapping signals. In Figure \ref{fig:noise} the difference between $S_{A \text{,residual}} (f)$ and $S_{A \text{,partial}} (f)$ is clearly visible for $f \in [\SI{2}{mHz}, \SI{10}{mHz}]$ where most signals are found.

\subsection{Pipeline}

To conclude we present in Algorithm \ref{alg:pipeline} the full pipeline to extract GBs within a given frequency range of $ f_\text{min}$ and $f_\text{max}$. The output is the list of MLEs $\tilde\theta_\text{recovered}$ and the list of MCMC chains $\tilde\theta_\text{posteriors}$ which provide the posterior distribution.

\begin{algorithm}[!ht]
\caption{The pipeline to obtain the MLE and posterior distribution of GBs within a large frequency range.}\label{alg:pipeline}
$\textbf{Function} \ \textbf{\textit{extracting\_GBs}}(f_\text{min}, f_\text{max}) $ \\
$B_\text{search} \gets \textbf{\textit{segmenting}}(f_\text{min}, f_\text{max})$ \\
split $B_\text{search}$ into $B_\text{even}$, $B_\text{odd}$ \\
$\tilde\theta_\text{even} \gets \textbf{\textit{global\_GB\_search}}(  B_\text{even}, 3, d)$\\
$\tilde\theta_\text{odd} \gets \textbf{\textit{global\_GB\_search}}(  B_\text{odd}, 10, d - \sum\limits_{\theta \in \tilde\theta_\textrm{even}} s(\theta))$\\
$\tilde\theta_\text{even} \gets \textbf{\textit{global\_GB\_search}}(  B_\text{even}, 10, d - \sum\limits_{\theta \in \tilde\theta_\textrm{odd}} s(\theta))$\\
$\tilde\theta_\textrm{recovered} \gets \tilde\theta_\textrm{even} \cup \tilde\theta_\textrm{odd}$\\
$\tilde\theta_\text{posteriors} \gets \{ \} $\\
$d_\text{residual} = d - \sum\limits_{\theta \in \tilde\theta_\text{recovered}} s(\theta)$ \\
$ \textbf{for all } \theta_\text{MLE} \textbf{ in } \tilde
\theta_\textrm{recovered} \textbf{ do in parallel} $\\
    \hspace*{1em} $d_\text{posterior} = d_\text{residual} + s(\theta_\text{MLE})$ \\
    \hspace*{1em} get $\Theta_\text{reduced}$ as described in Section \ref{sec:posterior}\\
    \hspace*{1em} $\tilde\theta_\text{posteriors} \text{ append } \textbf{\textit{posterior}}(\theta_\text{MLE}, d_\text{posterior}, \Theta_\text{reduced})$\\
\textbf{end for}\\
\textbf{return } $\tilde\theta_\text{recovered}$, $\tilde\theta_\text{posteriors}$\\
\end{algorithm}

\section{Results}
\label{sec:results}

The analysis of Radler's, LDC1-4, started with the first $\SI{0.5}{yr}$ and continued with $\SI{1}{yr}$ and $\SI{2}{yr}$, where the found signals from  the previous analysis are used as initial guesses for the DE algorithm. The Sangria data set, LDC2a, with the massive black hole binaries subtracted, is analyzed once for the full $\SI{1}{yr}$ of data. The global frequency band is set to $f_\text{min} = \SI{0.3}{mHz}$ and $f_\text{max} = f_\text{Nyquist}$ where $f_\text{Nyquist} = \SI{33.\overline{3}}{mHz}$ for the Radler challenge and $f_\text{Nyquist} = \SI{100}{mHz}$ for the Sangria challenge.

\subsection{Computation times}
\label{sec:computation}
Each segment of $f_\text{search}$ in $\textbf{\textit{global\_GB\_search}}$ can be analyzed in parallel as noted with $\text{"do in parallel"}$ in Algorithm \ref{alg:GB global}. Therefore the shortest time to analyze the data set $T_{\text{parallel}}$ is determined by the sum of the three frequency segments which take the longest for each sequential analysis of $B_\text{even}$, $B_\text{odd}$ and $B_\text{even}$ segments as listed in Table \ref{tab:search parameters}. The duration to analyze a segment varies a lot, where segments with no detectable signal are analyzed within $\SI{2}{min}$ and the longest computation time of a segment containing multiple detectable signals took $\SI{126}{min}$.

The data analysis to obtain the MLEs was run on a high-performance computer. In Table \ref{tab:evaluation time} we present the search times for finding the MLE solutions of the Radler data set. The pipeline demonstrates its efficiency by analyzing the longest observation time of $T_\text{obs} = \SI{2}{yr}$ in only $\SI{6}{h}$. In terms of computational cost, the analysis necessitates approximately $\SI{3\,300}{h}$ of CPU core hours. If commercial high-performance computing services such as those provided by Google are utilized, the estimated cost would amount to approximately $\SI{100}{USD}$ \cite{Google}.

\begin{table}[!htbp]
\caption{Computational times of the Radler LDC1-4 data with different $T_\text{obs}$. The CPU time is the sum of the computational time of all analyzed frequency segments. $T_\text{parallel}$ is the shortest computation time if the segments are analyzed in parallel on multiple CPU threads.}
\begin{ruledtabular}
\begin{tabular}{ccccc}
Challenge & $T_{\text{obs}}$ ($\si{yr}$) & CPU core time ($\si{h}$) & $T_{\text{parallel}}$ ($\si{h}$)\\ \hline
Radler & 0.5 & $1\,607$ & 3.2 \\
Radler & 1 & $2\,106$ & 4.3 \\
Radler & 2 & $3\,269$ & 5.5 \\
\end{tabular}
\label{tab:evaluation time}
\end{ruledtabular}
\end{table}

Furthermore, the computation of posterior distributions according to Section \ref{sec:posterior} takes 1.8 seconds per signal on a Quadro RTX 4000 Mobile GPU. Therefore, for example for the $8\,385$ recovered signals of the Sangria challenge it took $\SI{4.2}{h}$ on a single laptop to compute all posterior distributions.

\subsection{Matching recovered signals with injected signals}
\label{sec:matching}
To evaluate the accuracy of the recovered signals $\theta_\text{rec} \in \tilde\theta_\text{recovered}$, we are matching them with the injected signals $\theta_\text{inj} \in \tilde\theta_\text{injected}$ with similar frequencies.
In order to determine matches quantitatively, we use the scaled error

\begin{equation}
\delta(s(\theta_\text{rec}),s(\theta_\text{inj})) = \frac{ \langle  s(\theta_\text{rec})-s(\theta_\text{inj}),  s(\theta_\text{rec})-s(\theta_\text{inj}) \rangle} {\langle  s(\theta_\text{rec}), s(\theta_\text{rec}) \rangle}\label{eq:match_metric}
\end{equation}

which is dependent on the amplitude of the signals.

In other works the scaled correlation, also called overlap, 

\begin{equation}
\mathcal{O}(s(\theta_\text{rec}),s(\theta_\text{inj})) = \frac{ \langle  s(\theta_\text{rec})), s(\theta_\text{inj}) \rangle} { \sqrt{\langle  s(\theta_\text{rec}), s(\theta_\text{rec}) \rangle \langle  s(\theta_\text{inj}), s(\theta_\text{inj}) \rangle}}
\end{equation}

of two signals $s(\theta_\text{rec})$ and $s(\theta_\text{inj})$ is used \cite{damour2001comparison}.

\begin{figure}[!htbp]
\includegraphics[width=0.5\textwidth]{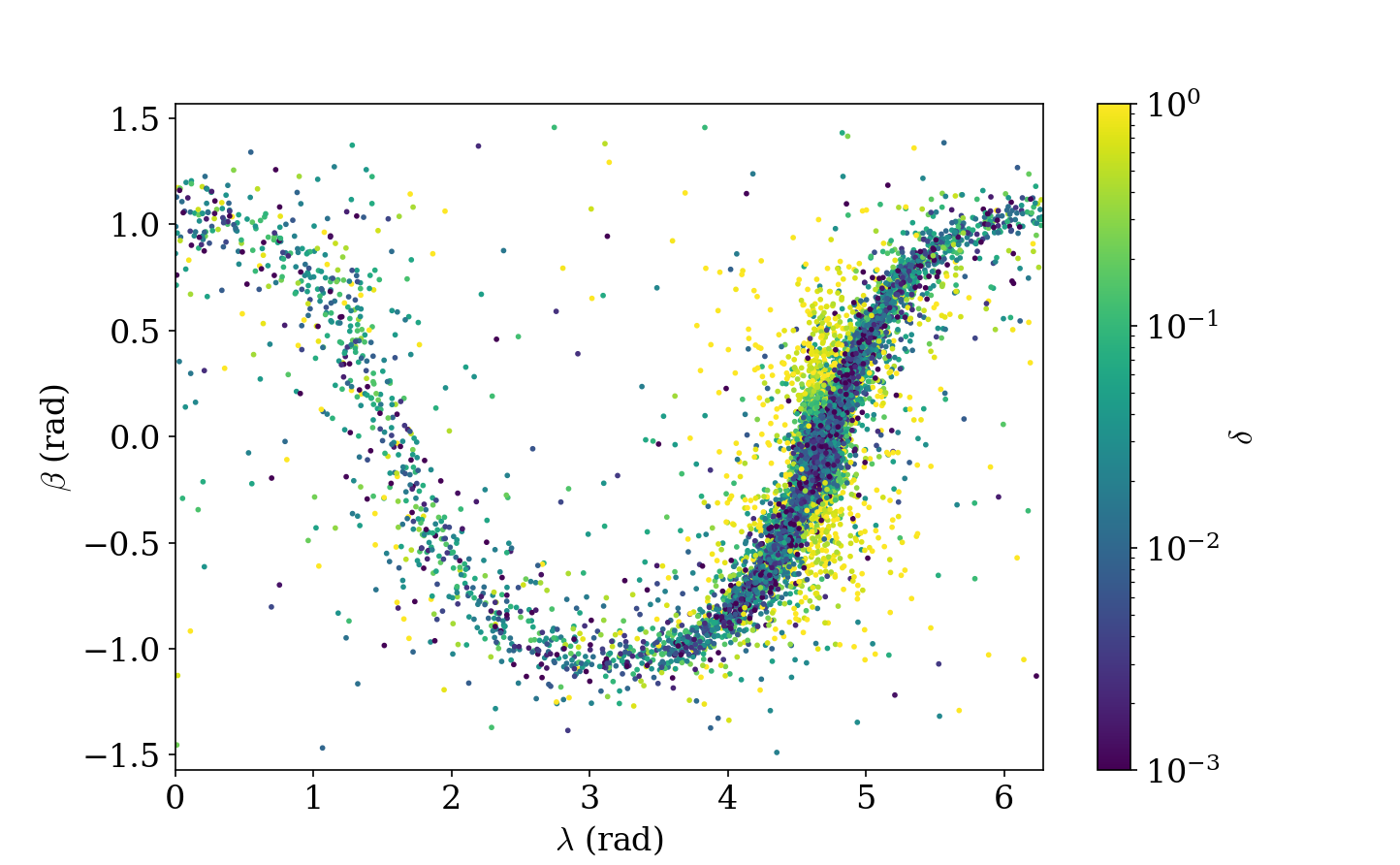}
\caption{\label{fig:skylocationerror} Scatter plot of the scaled error $\delta$ across the ecliptic longitude and ecliptic latitude of the $\SI{2}{yr}$ Radler data set. The range of the errorbar is clipped at $10^{-3}$ to $10^{0}$.}
\end{figure}

\begin{table*}[!htbp]
\begin{minipage}{\textwidth}
\caption{The variables of interest include the count of detectable injected GB sources, the count of recovered sources, the count of matches with injected sources, and the match rate. The match rate is determined by dividing the number of matched signals by the total number of recovered signals. The overlap is included to get an evaluation comparable to other analyses \cite{zhang2021resolving, gao2023fast, littenberg2023prototype}.}
\begin{ruledtabular}
\begin{tabular*}{\textwidth}{@{\extracolsep{\fill}}cccccccc@{\extracolsep{\fill}}}
Challenge     & $T_{\mathrm{obs}}$ [yr] & Injected ($\rho > 10$) & Recovered & $\delta < 0.3$ & Match rate $_{\delta < 0.3}$ & $\mathcal{O} > 0.9$ & Match rate $_{\mathcal{O} > 0.9}$\\ \hline
Radler     &  \num{0.5} & \num{6\,813} & \num{3\,937} & \num{3\,418} & \num{87 \%} & \num{3\,407} & \num{87 \%}\\
Radler     &  \num{1} & \num{11\,814} & \num{7\,112} & \num{6\,270} & \num{88 \%} & \num{6\,251} & \num{88 \%}\\
Sangria     &  \num{1} & \num{11\,814} & \num{8\,385} & \num{7\,173} & \num{86 \%}  & \num{7\,186} & \num{86 \%}\\
Radler     &  \num{2} & \num{18\,332} & \num{11\,952} & \num{10\,369} & \num{87 \%} & \num{10\,363} & \num{87 \%}\\
\end{tabular*}
\label{tab:ldc1-4-matches}
\end{ruledtabular}
\end{minipage}
\end{table*}

\begin{figure*}[!htbp]
\minipage{0.5\textwidth}
  \includegraphics[width=\linewidth]{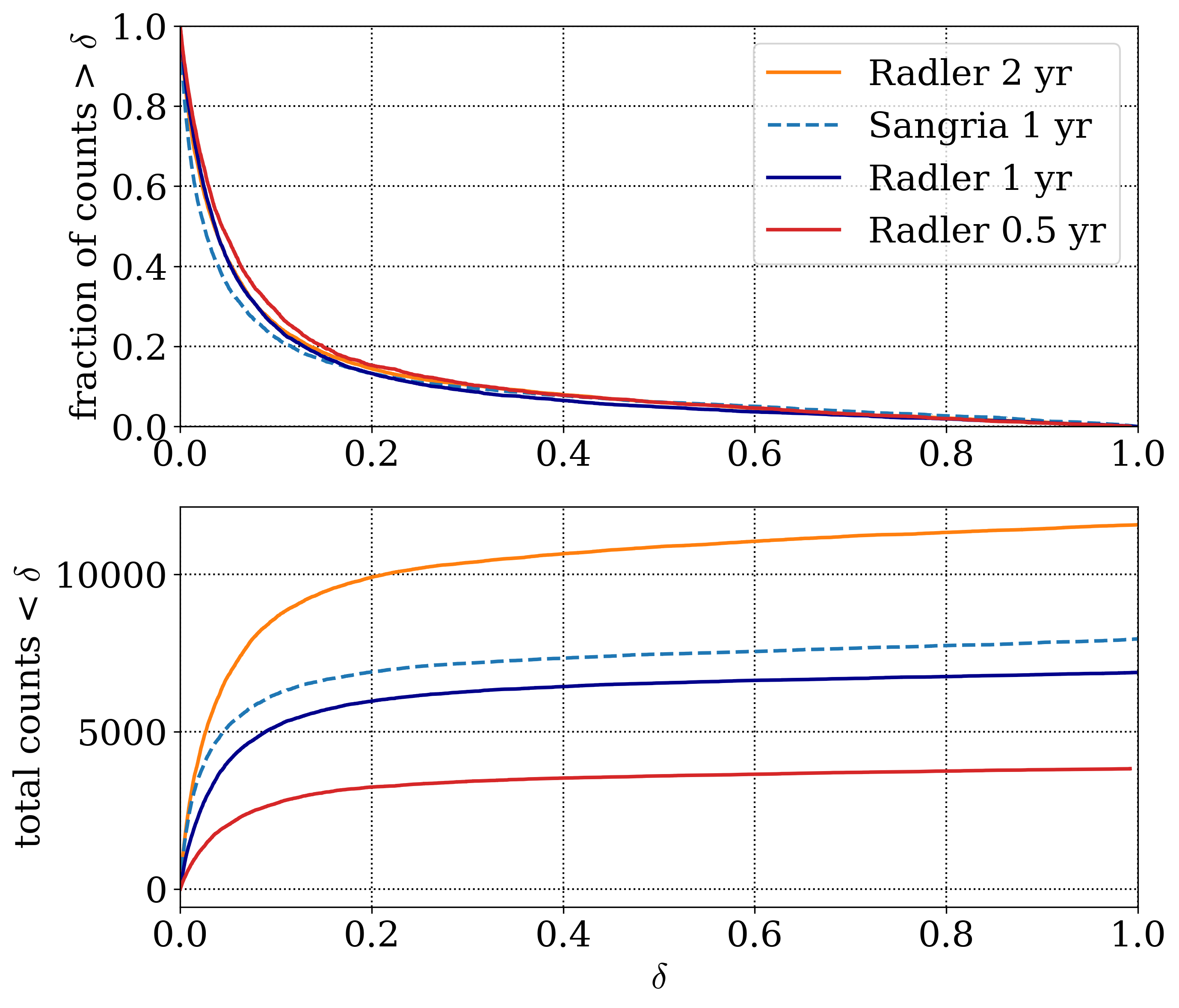}

\endminipage\hfill
\minipage{0.5\textwidth}
  \includegraphics[width=\linewidth]{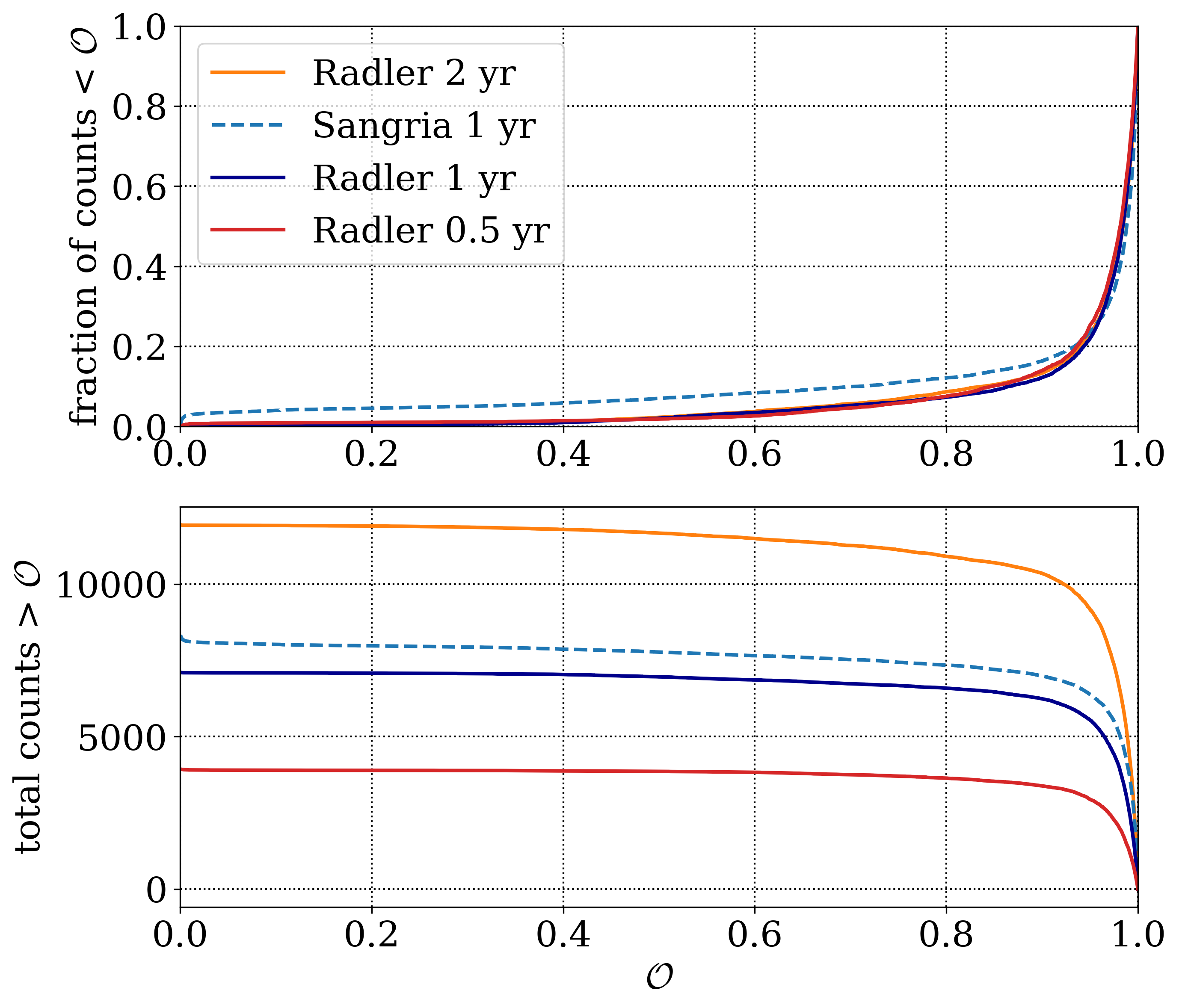}
\endminipage\hfill
\caption{Cumulative distribution function of $\delta$ and $\mathcal{O}$ in the top plots and the survival function in the bottom plots. The plots of the overlap $\mathcal{O}$, on the right, are comparable with other analyses such as \cite{littenberg2020global, littenberg2023prototype}}
\label{fig:cummulative}
\end{figure*}

\begin{figure*}[!htbp]
\includegraphics[width=1\textwidth]{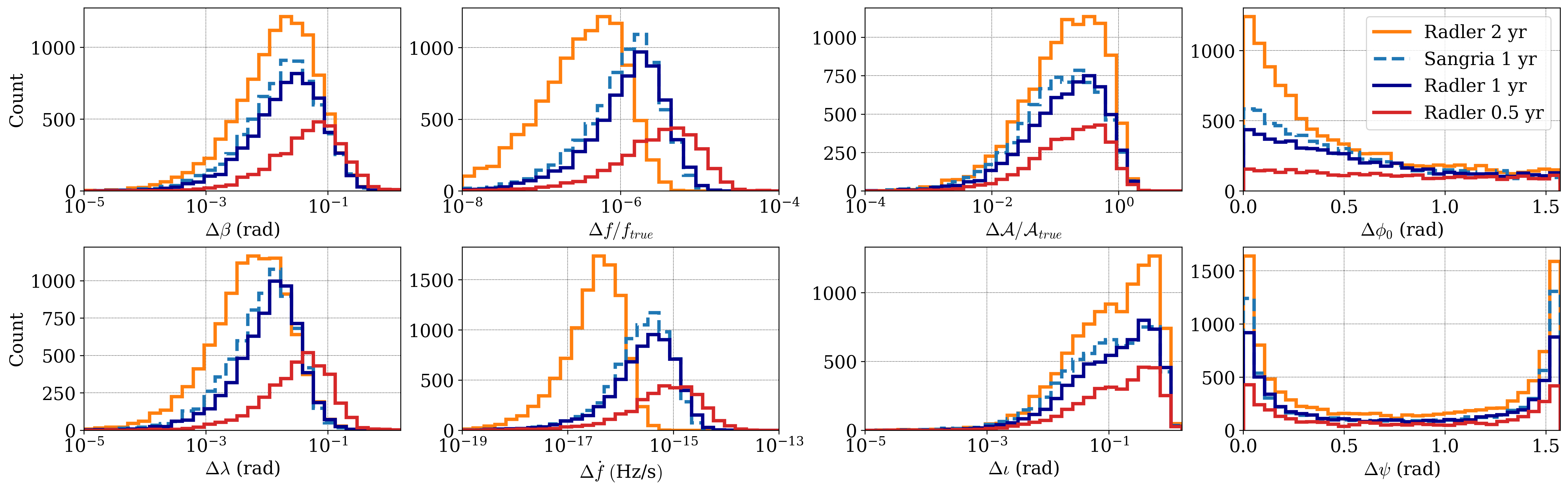}
\caption{\label{fig:error_histogram} Error histogram of all matched signals with $\delta < 0.3$.}
\end{figure*}
Figure \ref{fig:skylocationerror} shows the sky locations of all recovered signals in ecliptic coordinates. The recovered signals follow the geometry of the galaxy, with high $\delta$ (yellow dots), slightly off the center of the galaxy.
In Table \ref{tab:ldc1-4-matches} we present the number of recovered and matched signals for each analysis where we also include the overlap $\mathcal{O}$ as a match metric for comparison with other evaluations which used the overlap \cite{littenberg2020global, littenberg2023prototype, zhang2021resolving, gao2023fast}. The consistently high match rate of all analyses speaks of good quality recoveries. In Figure \ref{fig:cummulative} we see only small changes in the cumulative distribution function across the analyses. Only the cumulative distribution function of $\mathcal{O}$ for the Sangria data set has a higher count for smaller $\mathcal{O}$.

Given the potentially high correlation between a low-amplitude signal and a loud signal, even when the scaled error suggests a poor match, we classify the recovered signals with $\delta < 0.3$ as "matched" signals. For each matched signal, we calculate the error using $\Delta \beta = |\beta_\text{rec} - \beta_\text{inj}|$. In Figure \ref{fig:error_histogram}, we present the error histograms for all parameters. Notably, there is a clear trend of decreasing errors with longer observation times, as expected. The error histograms for the $\SI{1}{yr}$ analyses of the Radler and Sangria experiments exhibit similar patterns, consistent with our expectations. For the frequency and amplitude parameters, we display the relative errors. The relatively higher errors observed for $\phi_0$ and $\psi$ can be attributed to the inherent degeneracy between these two parameters. However, it is evident that the degeneracy diminishes with increasing $T_\text{obs}$.

\subsection{Galaxy}

The recovered signals that meet the matching criteria are visualized as green dots in Figure \ref{fig:amplitude scatter}. Additionally, it is evident from the plot that the recovered signals without a satisfactory match predominantly have lower amplitudes $\mathcal{A}$. The ability of LISA to recover signals is contingent upon the sensitivity curve, which exhibits lower sensitivity at lower frequencies. Consequently, only signals with higher amplitudes are recoverable at low frequencies. The majority of the recovered signals are concentrated in the central region of the Milky Way, which is also the location of a significant portion of the sources.

\begin{figure}[!htbp]
\minipage{0.5\textwidth}
\includegraphics[width=\textwidth]{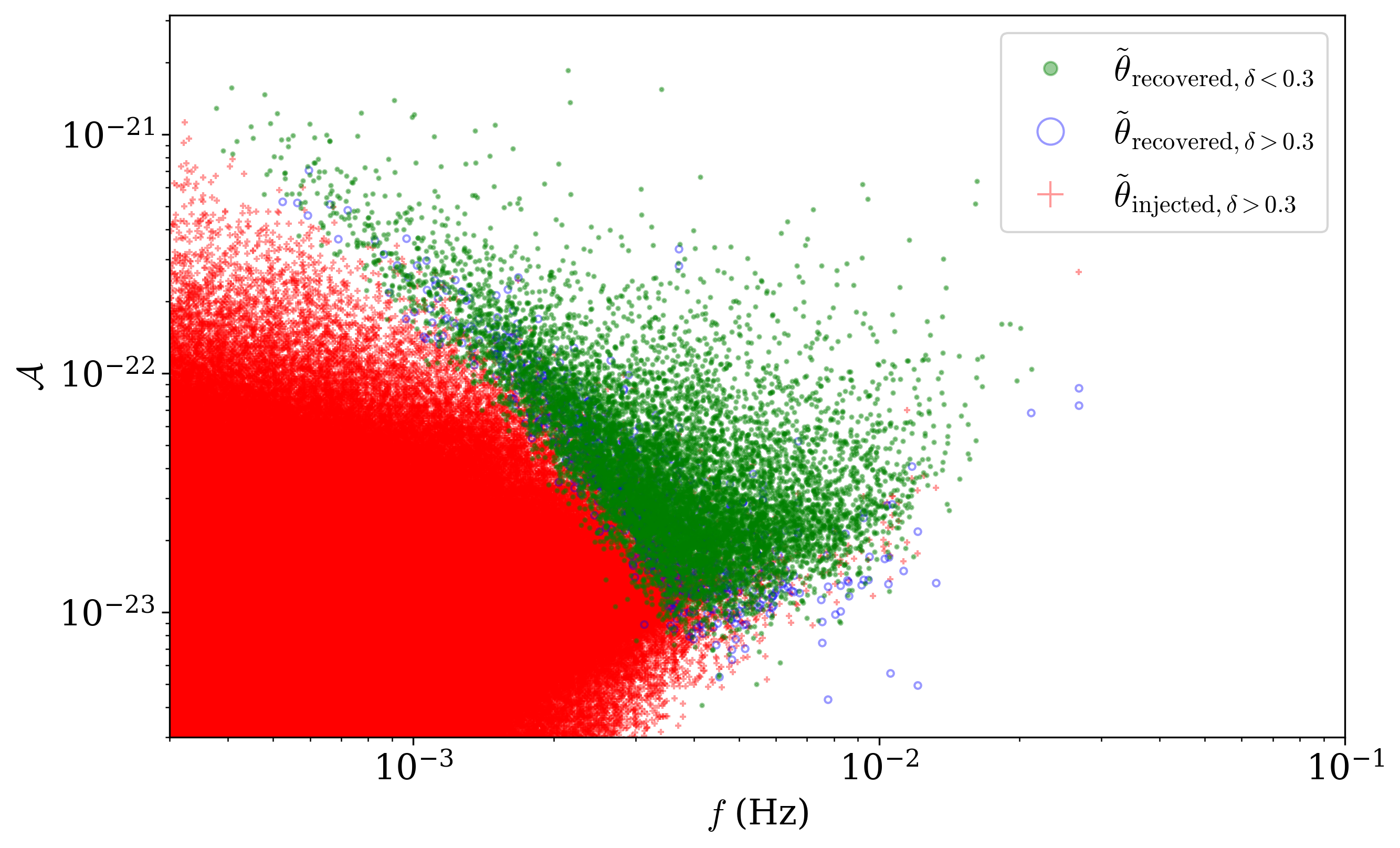}
\endminipage\vfill
\minipage{0.485\textwidth}
  \includegraphics[width=\linewidth]{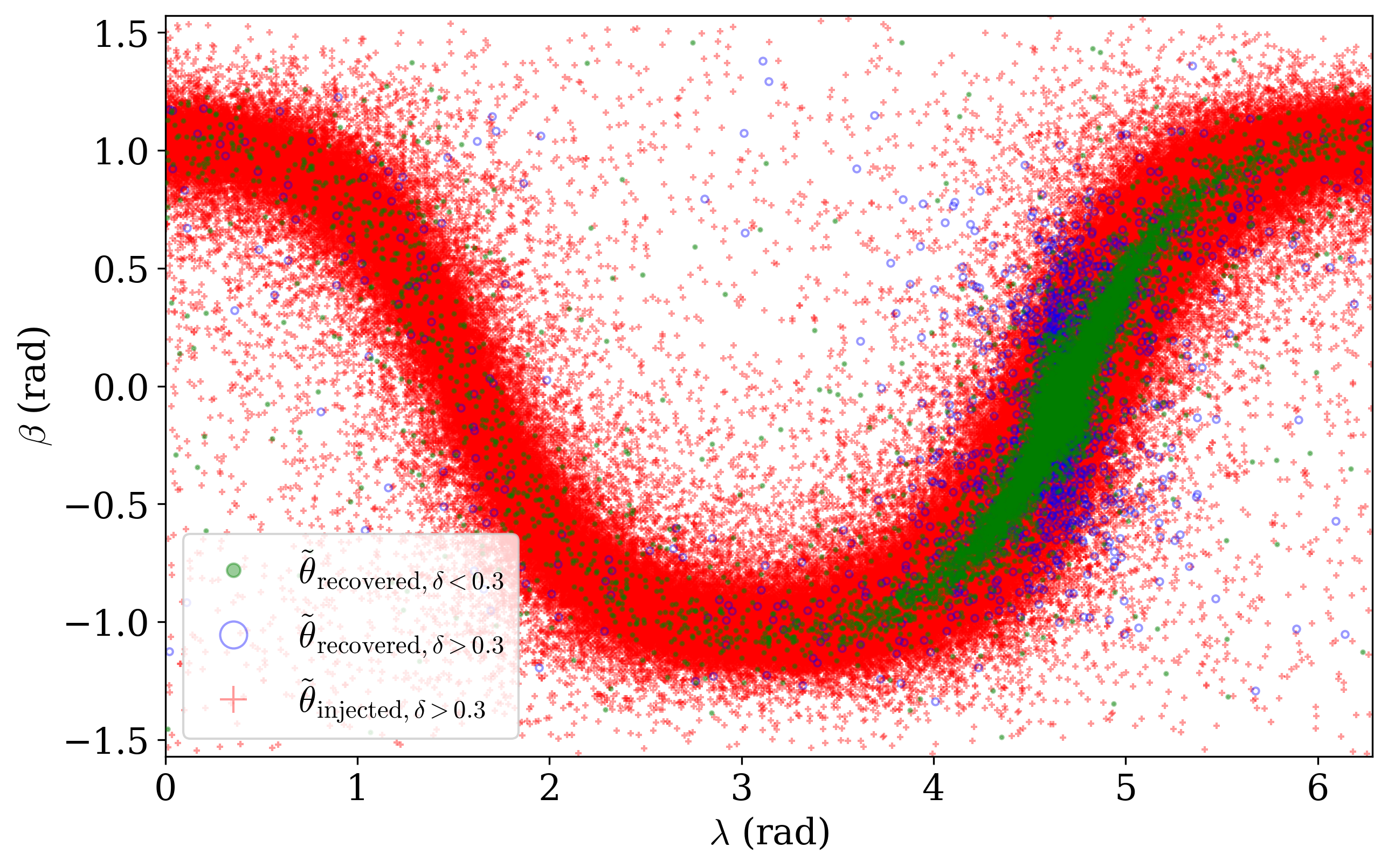}
\endminipage\vfill
\caption{\label{fig:amplitude scatter} Scatter plot of recovered GBs and injected GBs of the $\SI{2}{yr}$ Radler data set. The upper plot is across the GB amplitude $\mathcal{A}$ and frequency $f$ and the lower plot is across the ecliptic sky locations. The green dots are the recovered GBs with $\delta < 0.3$ which are categorized as matched signals. The blue circles on the plot represented the recovered signals that did not have a close match with any of the injected signals. The red crosses represent the injected signals that did not have a good match with any of the recovered signals. These signals were not effectively captured or identified during the recovery process.}
\label{fig:amplitude scatter}
\end{figure}

\begin{figure}[!htbp]
\minipage{0.5\textwidth}
  \includegraphics[width=0.9\linewidth]{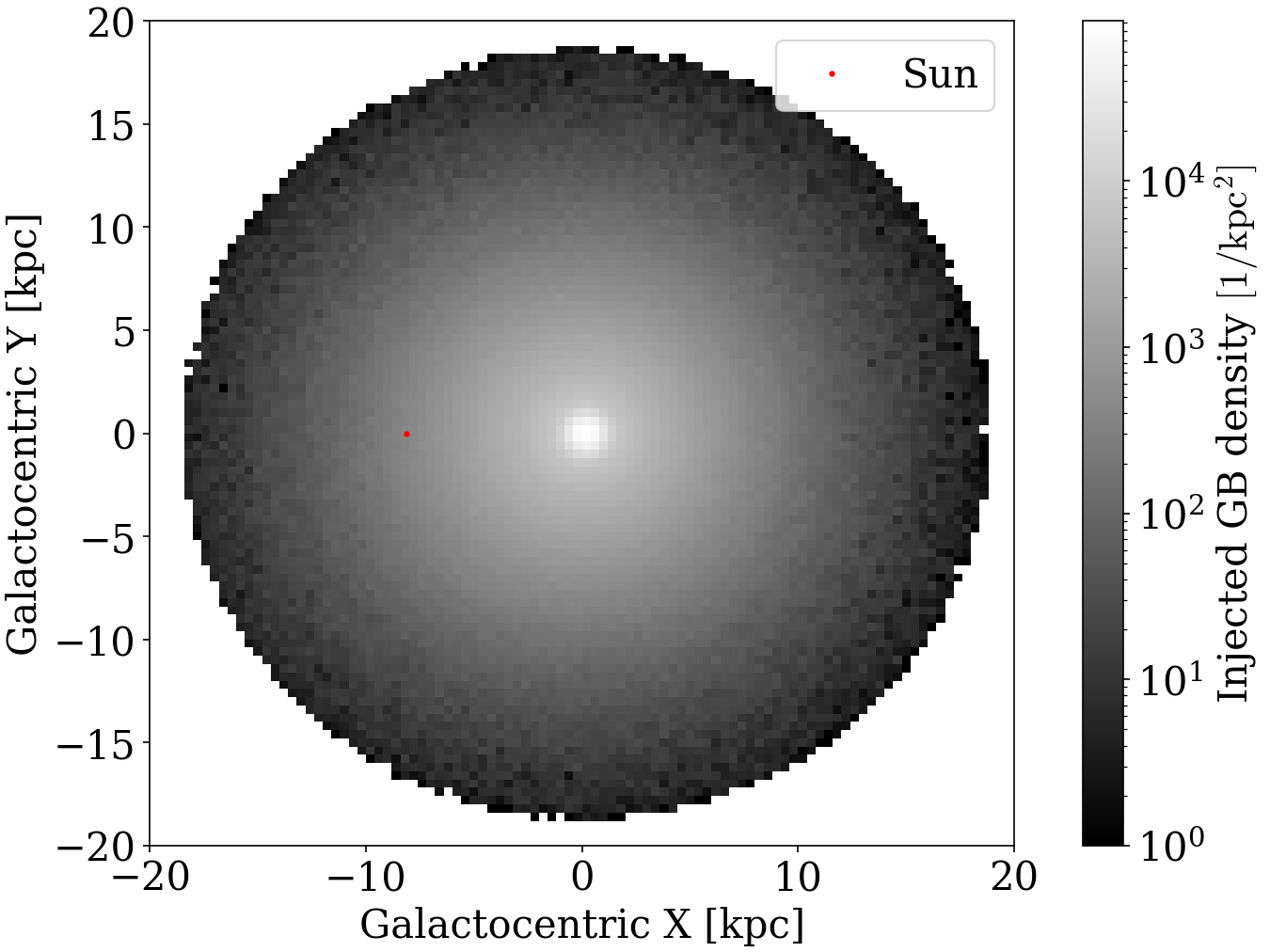}

\endminipage\vfill
\minipage{0.5\textwidth}
  \includegraphics[width=0.9\linewidth]{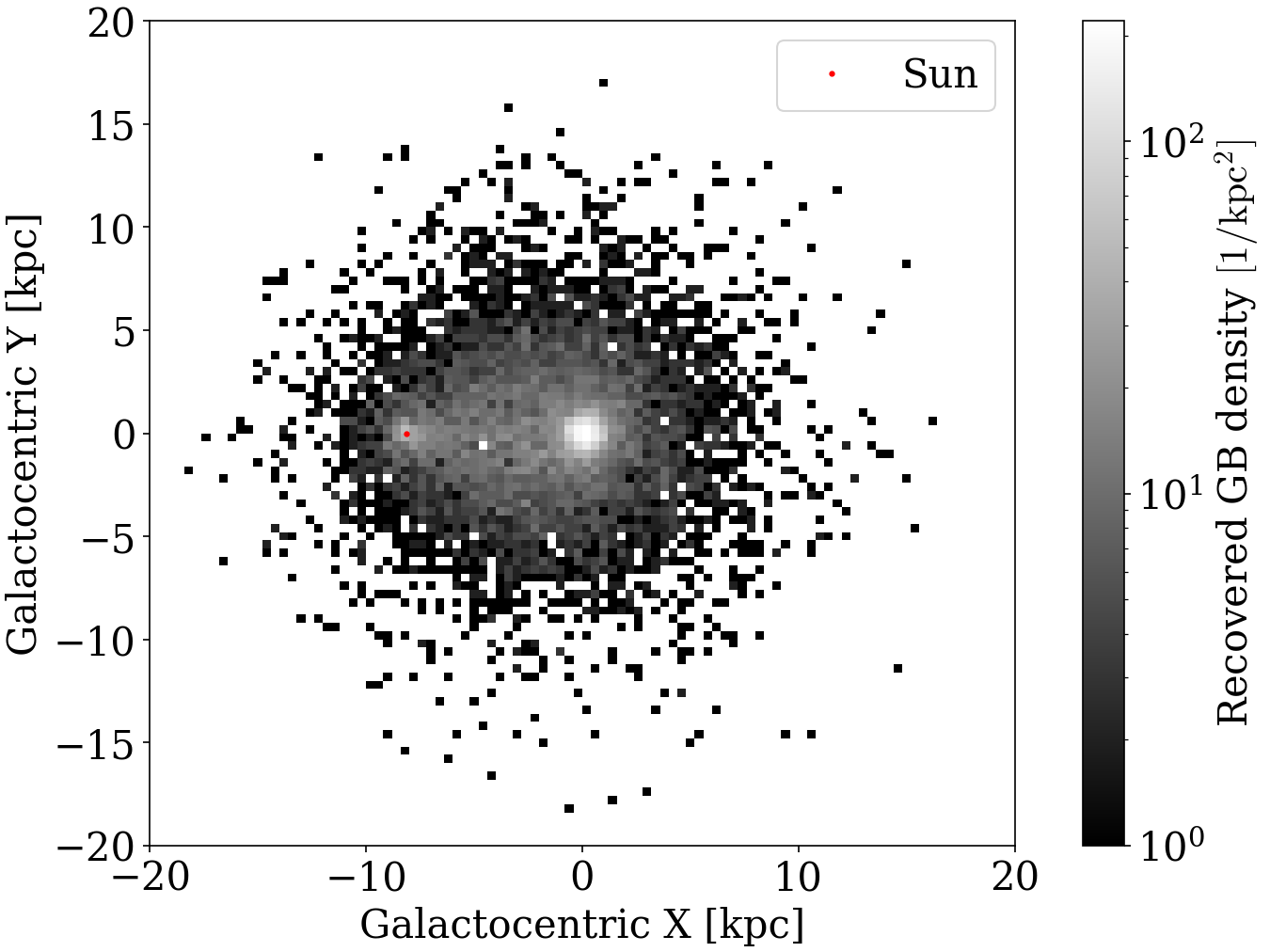}
\endminipage\vfill
\caption{The GB distribution of the Milky Way galaxy seen perpendicular to the galactic plane according to the simulated Radler data set. The red dot marks the sun. The top plot shows the distribution of the injected GBs and the bottom plot the distribution of the recovered GBs.}
\label{fig:galactocentric}
\end{figure}

\begin{figure}[!htbp]
\includegraphics[width=0.5\textwidth]{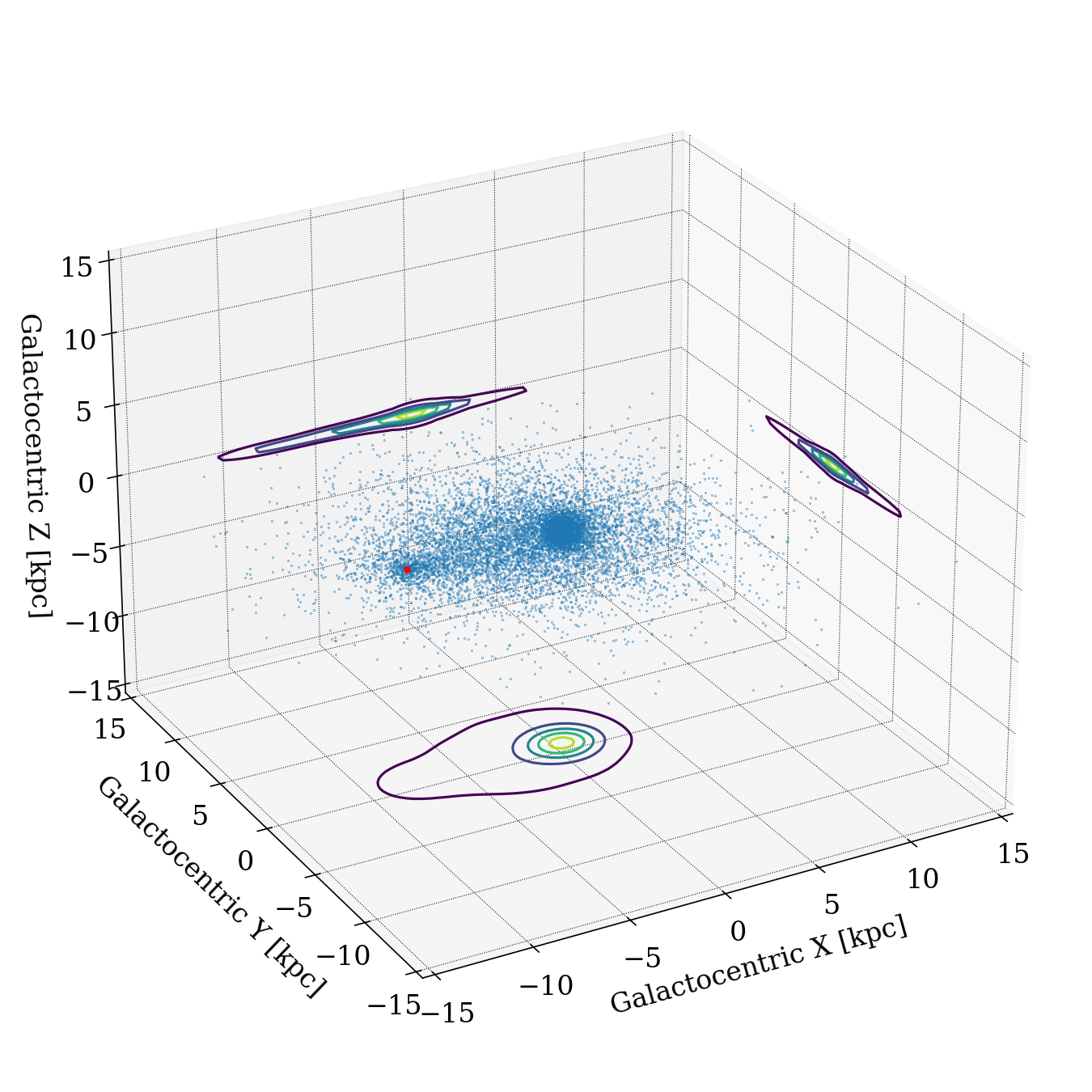}
\caption{\label{fig:galactocentric3d} Recovered GBs plotted as blue dots in the galactocentric coordinate system. The red dot marks the sun. The density of GBs is represented by 2D contour lines on the planes.}
\end{figure}

For each matched signal with $\dot{f} > 0$, where we assume that the evolution of the GB is purely driven by the emission of GWs, we can estimate the luminosity distance \cite{littenberg2020global}

\begin{equation}
D_L = \frac{5 \dot{f}}{48 \mathcal{A} \pi^{6/3} f_0^{5/3}}
\end{equation}

which is a good estimate of the distance in Euclidean space for objects in the Milky Way. Therefore we are able to convert their GBs to the galactocentric coordinate system and present them in Figure \ref{fig:galactocentric}. The upper plot illustrates the distribution of all injected signals with $f > \SI{0.3}{mHz}$ and $\dot{f} > 0$, while the lower plot depicts the recovered GBs.  It should be noted that the number of recovered GBs is lower than the injected ones due to the majority of injected GBs having a low (SNR), rendering them unrecoverable. Notably, a significant number of recovered GBs are located in close proximity to the sun, which aligns with expectations as closer sources exhibit higher SNR. This trend is also evident in the galactocentric 3D plot shown in Figure \ref{fig:galactocentric3d}.

\subsection{Posterior}
\label{sec:posterior quality}
\begin{figure}[!htbp]
\includegraphics[width=0.5\textwidth]{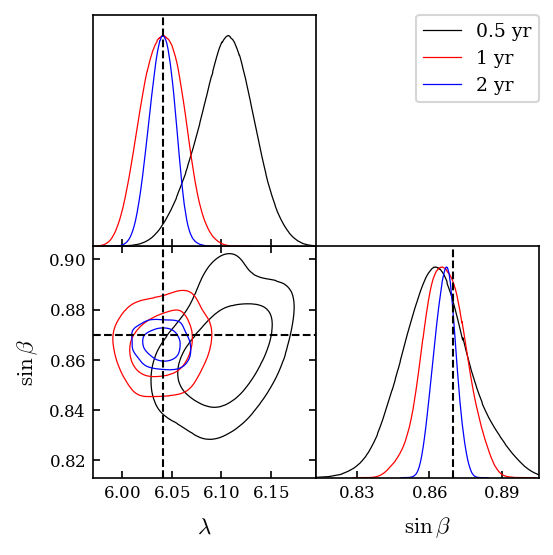}
\caption{\label{fig:posterior} Posterior distribution of the sky location for the 3 analyses of the Radler data set of the signal with $\theta^f_\text{inj} = \SI{4.169906}{mHz}$. The dashed black lines mark the true values of the matched injected signal $\theta_\text{inj}$.}
\end{figure}

Assessing the posterior distribution of $10\,000$ signals presents challenges, particularly in the absence of ground truth for comparison. However, leveraging statistical techniques allows us to evaluate the quality of the uncertainty estimates. Additionally, we can quantify the enhanced precision of the posterior distribution as $T_\text{obs}$ increases. In Figure \ref{fig:posterior}, we observe the evolution of accuracy and precision for one signal's sky location as a function of $T_\text{obs}$. Notably, for $T_\text{obs} = \SI{0.5}{yr}$, the accuracy and precision are comparatively lower than those achieved with longer $T_\text{obs}$.

\begin{figure}[!htbp]
\includegraphics[width=0.5\textwidth]{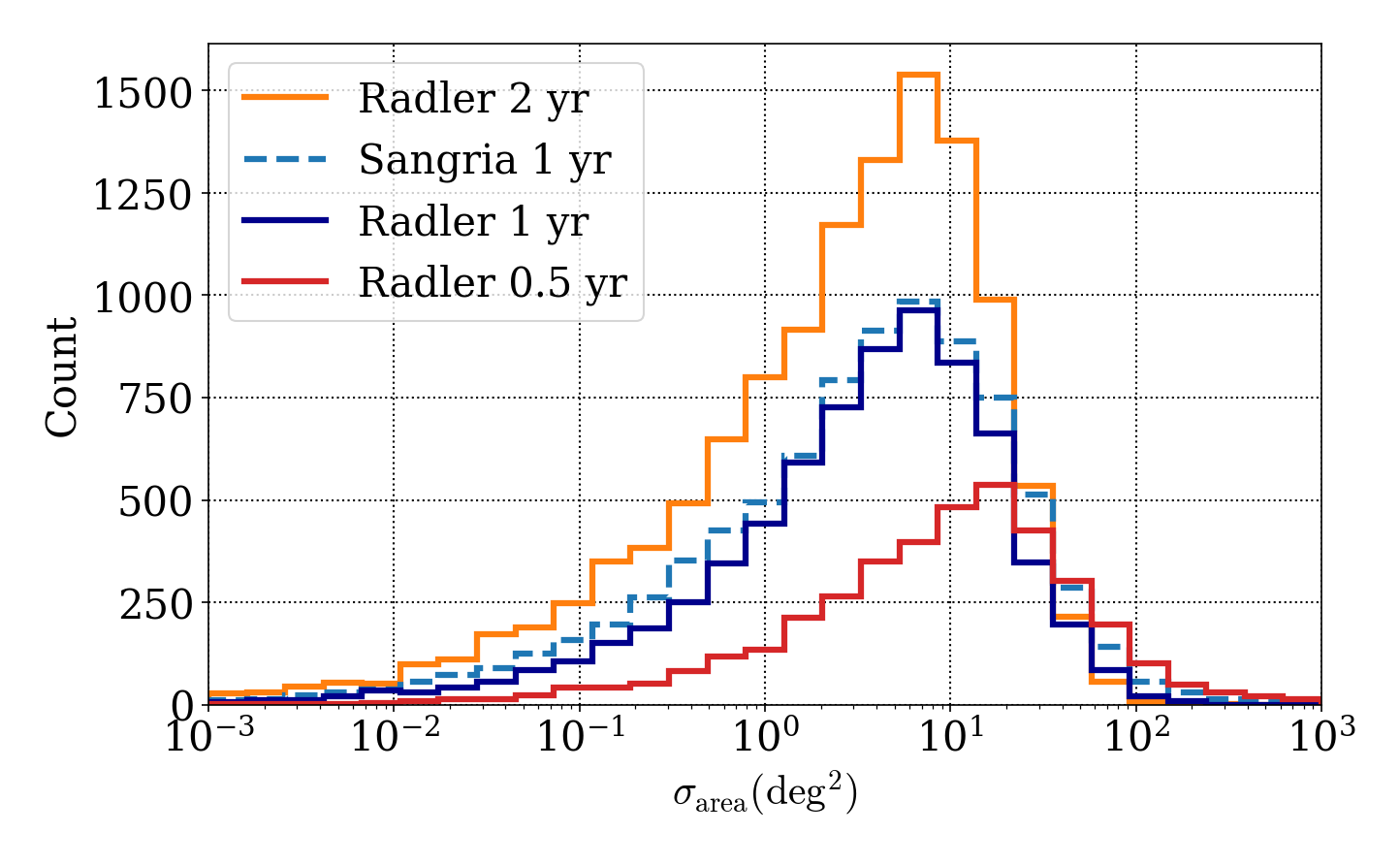}
\caption{\label{fig:angular} Angular confidence area histogram of all analyses.}
\end{figure}

Since the posterior distribution for the sky locations is approximate of Gaussian shape, we can estimate the uncertainty by computing the standard deviation $\sigma_\beta = \sqrt{ \frac{1}{N} \sum_{i=1}^N (\beta_i - \mu_\beta ) ^ 2}$ with mean $\mu_\beta = \frac{1}{N} \sum_{i=1}^N x_i$, where $N = n_\text{samples}$ is the length of the MCMC chain and $\beta_i$ is the $i^{th}$ sample of the chain. The uncertainty for the other parameters is computed analogously. In the next step, we can estimate the angular confidence area

\begin{equation}
    \sigma_\text{area} = \int_{\mu_\beta - \sigma_\beta}^{\mu_\beta + \sigma_\beta} \int_{\mu_\lambda - \sigma_\lambda}^{\mu_\lambda + \sigma_\lambda} \sin{\beta} \, d\lambda \, d\beta
\end{equation}

of the sky location for each signal. Figure \ref{fig:angular} displays the histogram of all analyses, revealing a notable trend. As $T_\text{obs}$ increases, the number of posteriors with small confidence areas also increases. This observation aligns with the findings depicted in Figure \ref{fig:posterior}, showing how the posterior of a signal becomes narrower with longer $T_\text{obs}$. However, it is important to note that the total number of extracted signals from the data also rises as the SNR of signals improves with longer observation times. Consequently, the number of signals with wider confidence areas also increases.

To assess the quality of the posterior estimate, we can examine whether the true parameters lie within the confidence interval as expected. If the accuracy and precision of the posterior are correct, we would expect to find the true parameters approximately $68\%$ of the time within the interval of $1 \sigma$ standard deviation.

If the number of parameters within the confidence interval is higher than expected, it suggests that the precision is worse, meaning the posterior distribution is too wide. On the other hand, if the number of parameters within the confidence interval is lower than expected, it indicates potential inaccuracies in the posterior estimate. This could mean that the posterior distribution is not located at the true parameters, and/or it is excessively precise, where the posterior distribution is too narrow.

\begin{table}[!htbp]\footnotesize
\caption{Ratio of true values within $1 \sigma$ standard deviation.}
\begin{ruledtabular}
\begin{tabular}{ccccc}
 Parameter &  Radler $\SI{0.5}{yr}$ & Radler $\SI{1}{yr}$ & Sangria $\SI{1}{yr}$ & Radler $\SI{2}{yr}$ \\ \hline
$\mathcal{A}$ & 23 \% & 29 \% & 34 \% & 33 \%  \\ 
$\sin \beta$ & 52 \% & 61 \% & 67 \% & 69 \%  \\
$\lambda$ & 24 \% & 54 \% & 60 \% & 72 \%  \\
$f$ & 20 \% & 28 \% & 32 \% & 39 \%  \\ 
$\dot{f}$ & 53 \% & 45 \% & 45 \% & 57 \%  \\ 
$\iota$ & 30 \% & 36 \% & 42 \% & 40 \%  \\ 
\end{tabular}
\label{tab:posterior accuracy}
\end{ruledtabular}
\end{table}

The results for individual parameters are presented in Table \ref{tab:posterior accuracy}, where we compute the standard deviation and check if the true parameter falls within the $68\%$ confidence interval. Due to degeneracy, the evaluation of $\phi_0$ and $\psi$ is omitted as it would not yield proper assessment. We observe that the uncertainty estimate for the sky locations, with observation times of $\SI{1}{yr}$ or more, is close to the expected rate. However, the other parameters exhibit a lower rate than expected. This can be attributed to multiple reasons. Firstly, it could be due to inaccurate estimation, where the true value does not align with the posterior distribution. Secondly, the posterior distribution might be too narrow. Lastly, the assumption of a Gaussian distribution for the other parameters, as used in computing the standard deviation, may not hold true. For multi-messenger astronomy, the good agreement between estimated and true uncertainty in sky location is of highest relevance.

\section{Conclusion}
\label{sec:conclusion}
The extention of the previous pipeline, outlined by \cite{strub2022}, allows now to obtain the MLE of GBs in the full frequency range where most of the GBs are overlapping with each other. As detailed in Section \ref{sec:computation}, the extraction of $18\,000$ signals from a data set with $T_\text{obs} = \SI{2}{yr}$ can be accomplished in a mere 6 hours. This acceleration also reduces the computational costs significantly to only 100 USD with today's hardware \cite{Google}, which brings extracting GBs from the full frequency band towards diminishing costs.

Additionally, we have leveraged the power of parallel computation, utilizing GPUs, to compute the posterior distribution for identified MLEs within a remarkable time frame of 2 seconds per signal. The computation of all posterior distributions can be completed in approximately 9 hours on a single laptop-grade GPU of the year 2018. These advancements not only enable efficient analysis of a large number of signals but also allow for rapid estimation of the posterior distributions.

The next crucial step involves integrating the presented pipeline into a comprehensive global analysis of data encompassing various astrophysical sources and phenomena, such as GBs, MBHBs, extreme mass ratio inspirals, glitches, and data gaps. The incorporation of this pipeline into the development of a global analysis framework offers substantial acceleration in the GB analysis process. This acceleration leads to a notable reduction in the associated costs for future pipeline developments.

\section{Acknowledgements}
We thank the LDC working group \cite{LDC} for the creation and support of the LDC1-4, LDC2a. Furthermore, we acknowledge the GPU implementation of $\texttt{FASTLISARESPONSE}$ \cite{michael_katz_2022_5867731}. The calculations of Algorithm \ref{alg:GB global} were run on the CPUs of the Euler cluster of ETH Zürich and are gratefully acknowledged. This project is supported by the Swiss National Science Foundation (SNF 200021\textunderscore185051). The full pipeline and evaluation tools are available at \cite{strub_stefan_2023_8123815}.

\bibliography{references}

\end{document}